%% file: main.tex
\documentclass{article}
\PassOptionsToPackage{authoryear,round}{natbib}
\usepackage[preprint]{neurips_2025}
\usepackage{graphicx} 
\usepackage[utf8]{inputenc} 
\usepackage[T1]{fontenc}    
\usepackage{hyperref}       
\hypersetup{hidelinks}
\usepackage{url}            
\usepackage{booktabs}       
\usepackage{amsfonts}       
\usepackage{nicefrac}       
\usepackage{microtype}      
\usepackage{xcolor} 
\usepackage{pdflscape}   
\usepackage{makecell}
\usepackage{longtable}
\usepackage{array}
\usepackage{enumitem}
\usepackage{xurl}
\renewcommand{\arraystretch}{1.2} 
\setlist[itemize]{nosep,leftmargin=*,topsep=0pt}

\usepackage[labelfont=bf]{caption} 
\usepackage[acronym]{glossaries}
\makeglossaries
\input{glossary.tex}
\setlist[itemize]{label=\textbullet}

\title{STAMP/STPA Informed Characterization of Factors Leading to Loss of Control in AI Systems}

\author{
Steve Barrett$^{1,*}$ \And
Anna Bruvere$^{1}$ \And
Sean~P. Fillingham$^{1}$ \And
Catherine Rhodes$^{1}$ \And
Stefano Vergani$^{1,2}$ \\
\\
$^{1}$ Arcadia Impact AI Governance Taskforce \\
$^{2}$ Department of Physics, King’s College London, UK \\
$^{*}$ Corresponding author: \texttt{sbarrett.work321@gmail.com}
}

\date{December 2025}

\begin{document}

\maketitle
\begin{center}
December 2025
\end{center}

\begin{abstract}
A major concern amongst AI safety practitioners is the possibility of loss of control, whereby humans lose the ability to exert control over increasingly advanced AI systems. 
The range of concerns is wide, spanning current day risks to future existential risks, and a range of loss of control pathways from rapid AI self-exfiltration scenarios to more gradual disempowerment scenarios. 
In this work we set out to firstly, provide a more structured framework for discussing and characterizing loss of control and secondly, to use this framework to assist those responsible for the safe operation of AI-containing socio-technical systems to identify causal factors leading to loss of control. 
We explore how these two needs can be better met by making use of a methodology developed within the safety-critical systems community known as \gls{stamp} and its associated hazard analysis technique of \gls{stpa}.
We select the STAMP methodology primarily because it is based around a world-view that socio-technical systems can be functionally modeled as control structures, and that safety issues arise when there is a loss of control in these structures. 
\end{abstract}

\newpage
\section*{Executive Summary}

\textbf{Why produce a framework for characterizing loss of control of AI systems?}
\begin{itemize}
\item There is a wide literature on the topic of loss of control and a correspondingly broad variety of conceptualizations of what loss of control means.  There would be benefit in providing a loss of control characterization framework that enables this wide variety of concerns to be conceptualized and related to each other and by which safety gaps can be systematically identified.    
\item Those charged with designing, deploying, and operating systems that contain AI can benefit from a framework that will enable them to better identify and mitigate the concern of loss of control.
\end{itemize}
\textbf{How does the topic of loss of control relate to safety?}
\begin{itemize}
\item According to the STAMP methodology, safety concerns arise when a system is no longer operating within specific bounds, known as safety constraints.  Many socio-technical systems can be functionally modelled as control systems, wherein a controller acts to ensure that the system always operates within its safety constraints.  When a system with such controls comes to operate outside of its safety constraints,  this is because control has been lost.  As such, the emergent property of safety is directly related to the concern of loss of control. 
\end{itemize}
 
\textbf{How can \gls{stamp}/\gls{stpa} be used to help fulfil the need for a framework for characterizing loss of control of AI systems?}
\begin{itemize}
\item STPA can be used to characterize various causal pathways by which loss of control can manifest.
\item The causal pathways by which loss of control can manifest in a generic control system can be linked to specific characteristics of AI, to thereby explain how loss of control may manifest in systems containing AI.
\end{itemize}

\textbf{Key contributions of this paper}
\begin{itemize}
\item Exploration of using \gls{stamp}/\gls{stpa} as a possible grounding for the characterization of causal pathways to AI loss of control.
\item Provision of a set of AI-related prompts, with associated methodology that supports identification of causal factors and pathways by which loss of control can manifest in AI systems.
\item This is distinct from other work \citep{mylius_systematic_2025} and \citep{ayvali_beyond_2025} that demonstrate the application of STPA to AI systems. 
\item Our exploration reported in this paper focuses on a simple, yet important and fundamental, control system archetype where we have considered the operating lifecycle phase.  As such, there are a number of important more complex potential AI loss of control scenarios and control system archetypes, which our work has not yet fully addressed.  These include, for example, systems with multiple AI agents, AI systems that recursively self-improve or modify their own training process or architecture, systems where no single controller exists and systems with hybrid human-AI controllers.
\end{itemize}

\newpage
\tableofcontents
\newpage
\section*{Contribution Statement}

Author Contributions (CRediT taxonomy)

\textbf{Conceptualization:} Steve Barrett, Ben R. Smith \\
\textbf{Methodology, Investigation:} All authors \\

\textbf{Writing – Original Draft, Review:} \\
\vspace{0.2cm}
\begin{tabular}{@{}ll@{}}
$\quad$ Steve Barrett      & Sections 1, 3, 4, 5, 7, 9, 10  \\
$\quad$ Anna Bruvere         & Sections 6.1, 8, App A, App B \\
$\quad$ Sean P. Filllingham       & Sections 2.3, 5, 6.2, 8, App A \\
$\quad$ Catherine Rhodes      & Sections 2.2, 6.3, App A, App B\\
$\quad$ Stefano Vergani       & Sections 2.1, 6.4, App A \\

\end{tabular}

\textbf{Writing – \LaTeX:} \\
\vspace{0.2cm}
$\quad$ Sean P. Fillingham, Stefano Vergani\\

\textbf{Writing – Review \& Editing, Supervision:} \\
\vspace{0.2cm}
$\quad$ Steve Barrett (Lead Editor) \\

\textbf{Project Administration:} Steve Barrett, Ben R. Smith
\vspace{8cm}
\section*{Acknowledgements}
We wish to thank Richard Mallah of the Centre for AI Risk Management and Alignment (CARMA), for providing the original research question on levels of loss of control and for consultation on some aspects of loss of control.  We would like to thank Simon Mylius for sharing his experiences with applying \gls{stpa} to an AI control problem, and for his insights about the type of framework that might be useful to those developing systems containing AI.  We are grateful to John Thomas for permission to use some pictures from the \gls{stpa} handbook.  Thanks to Francesca Gomez for sharing her perspectives on loss of control as a harm.  Thanks to Joe Rogero and Anna Katariina Wisakanto for review comments.  Any remaining errors are our own.  Finally, thanks to Ben R. Smith of Arcadia Impact for his overarching management of the programme under which this work was undertaken.

\newpage
\section{Introduction}
\label{sec:Intro}
The topic of loss of control has received much attention within the AI safety community. 
However, there are a wide variety of conceptualizations in the literature regarding the scope and meaning of loss of control. 
For example, for some researchers the focus of loss of control is on the sudden breakout of a powerful, misaligned \gls{agi} or \gls{asi} from an AI developer’s lab, whilst for others, the focus is on a disempowerment of humanity that arises very gradually through increasing societal dependence on AIs \citep{kulveit_gradual_2025, krook_when_2025, kasirzadeh_two_2025}.  For others, the concerns are with loss of control of current day AI-containing systems. 
With this work, we set out to take steps towards providing a more structured framework for conceptualizing loss of control as it applies to systems that include AI and humans.

In addition, this framework is intended to provide practical guidance to safety and risk management practitioners, to help them to identify AI-specific causal pathways that could result in loss of control in the AI systems for which they have responsibility.

The foundation for our framework is based around \gls{stamp} and its associated hazard analysis technique of \gls{stpa} \citep{leveson_engineering_2012}. 
This is a well-established methodology that has been used over recent decades by safety practitioners working in safety-critical industries. The STAMP methodology is based around a world-view that systems can be functionally modeled as control structures, and that safety issues arise when there is a loss of control in these structures.  By applying \gls{stpa}, we can also move beyond component-based accident models to consider hazards associated with the interactions between system components.   

Whilst the methodology has most frequently been discussed in the context of safety, it can be applied equally well to other categories of stakeholder loss, such as security. 

\gls{stamp}/\gls{stpa} provides a rigorous control-theoretic lens for analyzing loss of control in AI-containing socio-technical systems, including those where failures arise from governance,  organizational structure and human-AI interaction.  This work and the framework do not attempt to address loss of control concerns such as alignment.  Nor is the framework directed toward answering questions such as whether more capable systems should be trained or deployed; rather, its purpose is to structure reasoning about control conditional on such systems existing.  Whether the framework can be applied to \gls{agi} or \gls{asi} is an open question, and naturally there can be no current evidential support.

This paper is structured as follows: in Section~\ref{sec:OverviewPerspectives} we review the loss of control literature from a number of perspectives, including AI safety, existential risk, and safety-critical systems.

In Section~\ref{sec:selectSTPA} we identify the objectives for our work as regards this framework and provide a rationale for our selection of \gls{stpa} as the basis for our framework. 
In Section~\ref{sec:controlsystem} we provide an overview of some important functional control system archetypes that should be addressed by a more comprehensive characterisation framework, and describe the control system archetype that forms the focus for the remainder of our paper. 
In Section~\ref{sec:overview_LoCFramework} we provide a high-level overview of our framework and explain the details of how safety practitioners, with an interest in managing the risks of AI, can systematically make use of the framework to assess risk in their particular domain of interest.
In Section~\ref{sec:causalfactortable} we describe the method that was used to populate our characterization framework with AI-relevant control system insights. This method involves identifying the causal factors leading to loss of control by consideration, in turn, of each component within the functional control system.  
In Section~\ref{sec:controlsystemdegradations} we discuss the way in which gradual degradations in the control system can eventually lead to a loss of control event.
In Section~\ref{sec:validation} we demonstrate the use  of our proposed framework, by applying it to a specific example AI system: an AI-based national intelligence agency surveillance system. 
Section \ref{sec:futurework} provides recommendations for future work whilst Section~\ref{sec:conclusions} provides conclusions.  A glossary is provided in Section~\ref{sec:glossary}.

\section{Overview of Perspectives on Loss of Control}
\label{sec:OverviewPerspectives}
In this section we consider 3 perspectives on loss of control:
\begin{itemize}
    \item Perspectives on loss of control amongst AI safety practitioners (Section~\ref{subsec:perspectives_loc})
    \item Perspectives on loss of control within the existential risk literature (Section~\ref{subsec:perspectives_er})
    \item Perspectives on loss of control within safety-critical systems engineering (Section~\ref{subsec:perspectives_scs})
\end{itemize}

\subsection{Perspectives on Loss of Control amongst AI Safety Practitioners}\label{subsec:perspectives_loc}

\subsubsection{Definitions}

The EU AI Act’s Code of Practice for General-Purpose AI Models defines loss of control as “\textit{Risks from humans losing the ability to reliably direct, modify, or shut down a model. 
Such risks may emerge from misalignment with human intent or values, self-reasoning, self-replication, self-improvement, deception, resistance to goal modification, power-seeking behaviour, or autonomously creating or improving AI models or AI systems}.”\citep{noauthor_general-purpose_2025}.

The International AI Safety Report defines a loss of control scenario as “\textit{A scenario in which one or more general-purpose AI systems come to operate outside of anyone’s control, with no clear path to regaining control}” \citep{bengio_international_2025}.

\subsubsection{Literature}

\textbf{Loss of control is a harm in and of itself} \citep{gomez_we_2025}: Loss of control is considered a harm, alongside more conventional losses and harms such as financial loss or loss of life. Losing control over an AI is a psychological harm that needs to be prevented regardless of other consequences.  This is in contrast to the more commonly cited perspective that loss of control leads to a hazardous system state that may lead to a harm.  In this latter scenario, loss of control itself is not a harm but the possible consequences are. If a person’s well-being is dependent on the correct operation of a system, which humanity is no longer in control of, then this represents a hazardous system state.

Other perspectives focus more on the way loss of control can occur.

\textbf{Active vs passive loss of control} \citep{bengio_international_2025}: In active loss of control the AI system behaves in ways that actively undermine human control.  In passive loss of control people stop exercising meaningful oversight over AI systems for a variety of reasons. 

\textbf{Intentionality in loss of control} \citep{greenblatt_ai_2023, terekhov_adaptive_2025}: Loss of control may arise unintentionally or intentionally. If intentionally, this may occur with the assistance of humans or the AI may autonomously act to take control.

\textbf{Speed of loss of control, Accumulative vs Decisive} \citep{kulveit_gradual_2025, aguirre_control_2025, kasirzadeh_two_2025}: Loss of control could happen at a variety of speeds, it could be instantaneous or it could take years, for example, as is the case with gradual disempowerment. This is also sometimes described in terms of decisive loss of control, i.e. an overt and quick takeover versus accumulative loss of control, a gradual accumulation of AI induced threats and vulnerabilities that accrue until a trigger point is reached at which point control is lost. 

\textbf{Loss of control due to dangerous capabilities} \citep{bengio_superintelligent_2025, mitchell_fully_2025, grey_ai_2025, cohen_regulating_2024}:
Here the focus is on the dangerous capabilities of AI, that lead to the loss of control, where the suggested solution is often to control or limit those capabilities.

Other literature focuses on how to analyze loss of control risk.

\textbf{Assessing risk of loss of control} \citep{wisakanto_adapting_2025, mylius_systematic_2025, corsi_considerations_2024, raza_trism_2025, bryan_taxonomy_nodate}:
Some literature argues that classic component failure or probabilistic frameworks cannot apply to loss of control in complex systems that include AI, and new approaches such as STPA, novel taxonomic risk frameworks, or new forms of probabilistic risk assessment are needed 

\textbf{Control inversion} \citep{aguirre_control_2025}: 
Frames loss of control as a control inversion, in which rather than human controllers being in control of the AI, the AI comes to control the human controllers.

There is then a whole literature on mitigating loss of control, for example, with works such as \citep{tegmark_provably_2023, dalrymple_towards_2024} that claim the need for provably safe AI in order to protect against loss of control, and \citep{bengio_superintelligent_2025} recommending an alternative approach to training AI agents.

\subsection{Perspectives on Loss of Control within Existential Risk Literature}
\label{subsec:perspectives_er}
Frontier AI loss of control opens up pathways to catastrophic and existential risks – the latter threatening the existence of humanity as a whole, durable disempowerment of humanity, or civilizational collapse so severe we cannot recover.

Highly capable AI systems over which human control has been lost could, through pursuit of goals unaligned with human well-being and survival, threaten our existence, controlling critical systems, diverting essential resources, and otherwise acting in ways directly or indirectly harmful to humans and to the physical and natural systems on which we depend. Pathways toward catastrophic and existential risk scenarios include through deliberate deployment by humans of highly capable AI models in specific critical systems and subsequent loss of control of those systems (e.g. nuclear command and control, power grids, financial systems), and through its self-integration in and propagation through a range of systems in pursuit of instrumental goals. Specific areas of concern identified in the literature include:
\begin{itemize}

\item Recursive self-improvement with associated ‘intelligence explosion’ \citep{booth_biggest_2025}.

\item Human inability to adequately understand and effectively intervene in systems of significantly advanced intelligence \citep{aguirre_control_2025}.

\item Socio-technical, economic and geopolitical drivers of rapid development and deployment of increasingly advanced AI systems \citep{aguirre_control_2025}.
\end{itemize}
There is a broad and expanding literature on existential and global catastrophic risks and their management. 
Some of this addresses specific sources or types of risk, and some of it looks across risks of this class, and, or takes a more systemic approach to their assessment and management \citep{jehn_state_2025}. 
This level of risk has also increasingly come to the attention of policy makers and governance organisations in recent years.

Risks from frontier AI have been a major focus for work on existential risk \citep{o_heigeartaigh_extinction_2025, jehn_state_2025} with loss of control forming a clear pathway toward outcomes of global catastrophic or existential severity. In this sense, much of the literature on AI loss of control (see section \ref{subsec:perspectives_loc}) might be considered part of the existential risk literature.

Regarding pathways of existential risks,  the 2018 paper ‘Classifying Global Catastrophic Risks’ outlines a multidimensional approach to this task. It develops an analytical framework with three dimensions: “the critical systems affected, global spread mechanisms, and prevention and mitigation failures”, demonstrating commonalities among various extreme risk scenarios \citep{avin_classifying_2018}. 
It includes a hierarchical perspective, for example, higher-level critical systems, such as ecological and socio-technical systems, relying on the function of lower-level systems such as physical and biogeochemical systems.

Relevant to consideration of how to characterize loss of control scenarios, ‘Classifying Global Catastrophic Risks’ points to the usefulness of understanding how frontier AI loss of control may impact critical systems, how those impacts might propagate, and how prevention and mitigation efforts might fail – ultimately breaching thresholds that threaten humanity’s survival. This implies that to effectively inform management of loss of control risks, we not only need to consider dimensions of the AI system(s), but also of the system(s) in which it is deployed, and our capabilities to intervene in them.

While the case studies we address in this paper represent loss of control of a control system of which AI is a component, rather than general loss of control of frontier AI, the analysis points toward characteristics likely to contribute to broader loss of control and / or loss of control in critical systems, and indicates some gradations of associated capabilities.

\subsection{Perspectives on Loss of Control within Safety-Critical Systems Engineering}
\label{subsec:perspectives_scs}
Traditional hazard analysis methods have been particularly successful when the failures of the system can be traced to broken or malfunctioning components. In these systems most of the behavior is deterministic and systems do not learn, adapt, or develop new capabilities. Many of the objectives in the system are explicitly programmed and stable under normal operating conditions. AI models themselves, and when they are deployed as part of a larger system may violate many of these conditions, thereby making many traditional hazard analysis techniques insufficient for analysing AI. 

AI systems may have a dual-role inside these systems, as both the controller and a controlled process. When AI is used in a controlled process it will be developed and trained by humans with certain organizational policies and technical limitations in mind. During deployment it will be monitored with some sort of “human in the loop”.  When AI is used in the controller, the AI may make autonomous decisions based on inputs and observations of the environment and pursue objectives that are consistent with its learned goals. The control systems that are integrated with AI systems are often hierarchical and can lead to risk even when every component is behaving exactly as intended. While many traditional hazard analysis processes may fail to identify the risks in these systems, \gls{stpa} provides a more appropriate framework to analyze loss of control risk in AI enabled systems.

\subsubsection{Use of \gls{stamp}/\gls{stpa} to Analyze Loss of Control}

System-Theoretic Process Analysis (\gls{stpa}) is a hazard analysis technique that is based on STAMP.  It can generally be applied through four steps that are briefly explained below and outlined in Figure 1; see \citep{leveson_stpa_2018} for a complete description of how to apply STPA to a complex system.

\begin{figure}[h]
  \centering
  \includegraphics[scale=0.55]{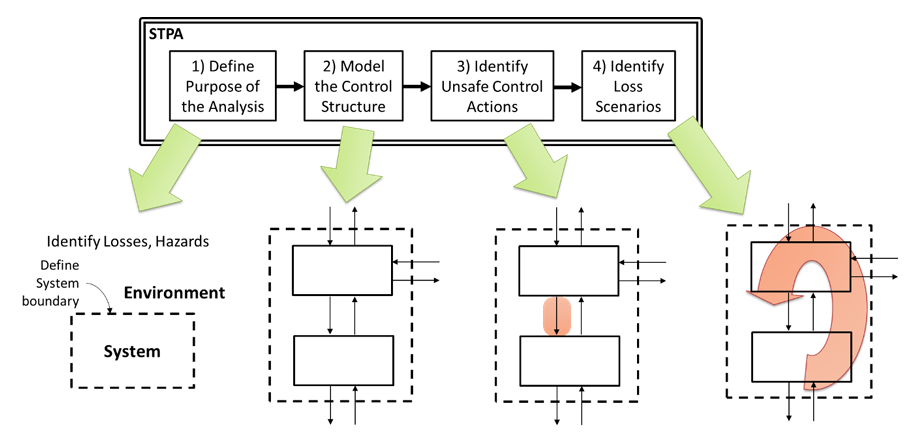}
  \caption{Overview of the high-level \gls{stpa} process, source: \gls{stpa} Handbook, thanks to John Thomas for permission to use this picture}
  \label{fig:Figure5.1}
\end{figure}

First, the losses and hazards need to be identified. The losses are the real-world outcomes that we are trying to prevent. They can take the form of existential harms to humanity (extinction), large-scale harms that fall short of extinction (economic collapse, authoritarian control), or any other harm that is undesirable. The hazards are states inside the system that can lead to losses. The specific nature of the hazards will depend on the context of the control structure and system design details. A critical insight is that each AI might be operating exactly as it was trained and still lead to hazards that may propagate into losses without proper controls and safety constraints.

The second major step in \gls{stpa} is to fully model the control structure and how each component interacts with the rest of the system. Figure~\ref{fig:Figure4.2} below provides an overview of how there can be controllers and controlled processes at multiple levels inside the control structure. The control actions (actuators) and feedback processes (sensors) help the controller constrain and understand the state of the controlled process. The exact details of this overall structure will depend on the system in question and will need to be carefully modelled in order to capture all of the potential hazards.

Using this model of the control structure, the third step identifies the various control actions and corresponding \gls{uca}s. There are four general criteria to consider when characterizing a \gls{uca}: 

\begin{itemize}
    \item Failure to provide the control action when necessary.
    \item Providing the control action when unnecessary.
    \item Implementing the control action at the wrong time (delay).
    \item Maintaining the control action for the wrong amount of time (duration). 
\end{itemize}

While all of these types of \gls{uca}s are of concern from a loss of control perspective, the wrong timing and wrong duration criteria could be particularly challenging to overcome due to the differences in speed that can sometimes arise between human controllers and AI.

The final step is to identify loss scenarios which describe the reasons \gls{uca}s occur, and the causal factors we can trace through the control structure for a particular \gls{uca}. The exact nature of these loss scenarios depends on the specifics of a particular loss, hazard, and \gls{uca} in the system. 

\gls{stpa} is a potentially powerful analysis tool for AI since it captures the systemic and complex nature of AI-embedded systems. \gls{stpa} can also identify risky scenarios that are not obvious. 
For example, there can be cases where every component in the control loop is operating properly but the environmental conditions and the state of the system can still lead to an unsafe control action and ultimately a loss. 
By applying the \gls{stpa} process, system constraints can be developed to mitigate the identified \gls{uca}s and hazards, helping to reduce the risks associated with the system.

\section{Selection of STPA as a Basis for a Loss of Control Characterization Framework}
\label{sec:selectSTPA}

\subsection{Objectives for our Loss of Control Framework}

\begin{itemize}
    \item[] \textbf{Objective 1}:  Development of a framework that provides the AI safety community with a more structured approach for discussing loss of control.
    \item[] \textbf{Objective 2}:  Development of a framework that can be used to provide practical guidance to individuals responsible for ensuring the safety of systems that incorporate AI, helping them to perform hazard analysis when applying a loss of control lens to their particular risk domains of interest. 
\end{itemize}

The work presented in this paper is conducted so as to advance progress towards meeting these objectives, by considering the merits of an approach informed by \gls{stamp}/\gls{stpa}.

\subsection{Rationale for selecting \gls{stamp}/\gls{stpa}}
\gls{stamp}/\gls{stpa} provides a useful framing for considering safety issues arising from loss of control in socio-technical systems \citep{leveson_engineering_2012}. The systematic approach of \gls{stpa} hazard analysis has been proven, in non-AI based socio-technical systems, to be capable of identifying hazards that were not identified via other common, and less structured, hazard analysis methods \citep{leveson_engineering_2012}. \gls{stpa} also provides the basis for a taxonomy of causal factors in loss of control, as can be derived from the figures in \citep{leveson_stpa_2018}.

In considering whether STPA might form a good basis for identifying and characterizing causal factors leading to loss of control of systems that include AI, it is useful to compare the types of causal factor which are suspected to lead to loss of control of AI systems, with the types of causal factor that \gls{stpa} is known to deal with and work for. To this end, it can be noted that \gls{stpa} is specifically designed to handle socio-technical systems \citep{leveson_engineering_2012}, in which humans may form parts of the control system, either in the controller, the controlled process, the sensing, or the actuation.   Humans and AI share some similarities, both are ‘grown’ and both make use of very large neural networks that are somewhat inscrutable black boxes.  Both may be situationally aware and exhibit undesirable, deceptive, or even malign behaviors. STPA has been applied to some systems where a weaker intelligence (e.g. a CEO) controls a super-intelligence (an organization of thousands of individuals), through hierarchical control structures. Of course, an AI component of a system is also different to a human ‘component’ of a system in many important respects.  For example, the speed, optimization pressure, and scalability of deception may be qualitatively different between humans and AIs. Control of an AI super-intelligence is also a different challenge to control of a human organization super-intelligence.  

In this paper we explore the potential of using \gls{stpa} as a basis for framing concerns around AI loss of control.  Of course, STAMP/STPA is just one approach amongst many for identifying hazards and addressing safety concerns.  We did not contrast a framework built on STAMP/STPA with frameworks built on other possible safety enabling world-views and methodologies. We make no claim that our framework is the best possible framework. We also observe that to our knowledge, there can be no guarantee with any hazard analysis technique (including \gls{stpa}) that all hazards and pathways to loss of control will be comprehensively identified.

That \gls{stpa} can prove valuable in identifying loss of control hazards in at least some types of system that include AI is demonstrated by our study results ( for a very simple control system archetype) that are presented in Sections~\ref{sec:overview_LoCFramework}, \ref{sec:causalfactortable} and Appendices~\ref{sec:appendixA} and \ref{sec:appendixB}. It is also demonstrated in work by \citep{mylius_systematic_2025}, where \gls{stpa} was applied to an AI system, and where it was found that \gls{stpa} can be used to identify causal factors that may be missed by unstructured hazard analysis methodologies.  The application of STPA during the development of AI has also been demonstrated in \citep{ayvali_beyond_2025}. 

There is a question around the range of applicability of this framework.  If, for example, it is the case that an 
\gls{agi}/\gls{asi} may be theoretically or practically uncontrollable, no effective control structure could be built. If no effective control structure can be built, this may raise the question of whether there is value in modeling the world in such a way.  One response here is that there can be value in such modeling to help designers understand that the system they are considering building is uncontrollable (and hence that it should not be built).  Nevertheless, we consider whether an \gls{stpa}-inspired taxonomy can be used to characterize all types of possible AI loss of control causal pathways and types of causal factor, including those associated with very advanced AIs to be a topic for future research.

\section{Overview of Functional Control System Archetypes}
\label{sec:controlsystem}
As outlined above, stage 2 of the STPA process is to model the functional control structure.  There are multiple functional control system archetypes that may be of interest in systems that make use of AI.  In this section we briefly present a likely non-exhaustive list of some of the key archetypes. 

\begin{figure}[h]
  \centering
  \includegraphics[scale=0.25]{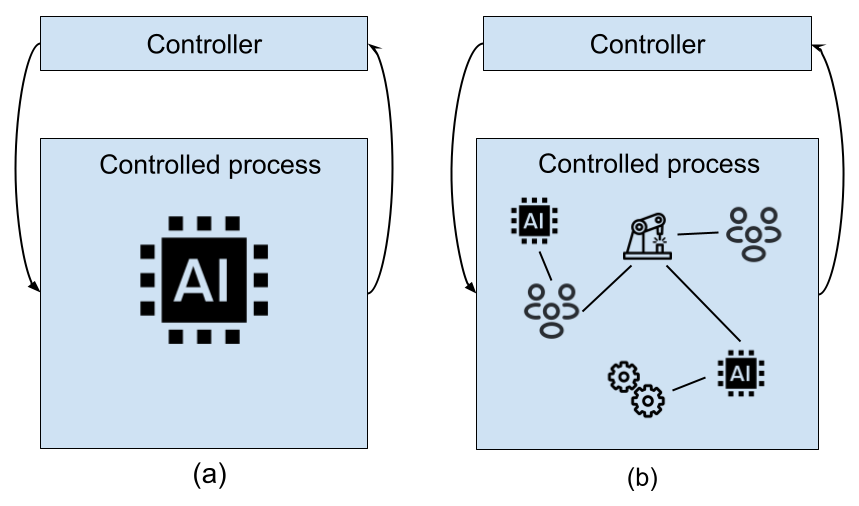}
  \caption{(a) Direct control of AI, (b) Control of a system that includes AI}
  \label{fig:Figure4.1}
\end{figure}

Figure~\ref{fig:Figure4.1} illustrates that control may be of the AI itself or a system that contains AI(s). 

\begin{figure}[h]
  \centering
  \includegraphics[scale=0.6]{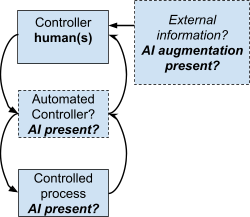}
  \caption{Control system archetypes with additional AI instantiations.}
  \label{fig:Figure4.2}
\end{figure}

Figure~\ref{fig:Figure4.2} illustrates control system archetypes where either a) an automated controller, which may or may not contain an AI, is used by the human(s) to control the controlled process, or b) human(s) use AI augmentation in the controller. 

\begin{figure}[h]
  \centering
  \includegraphics[scale=0.7]{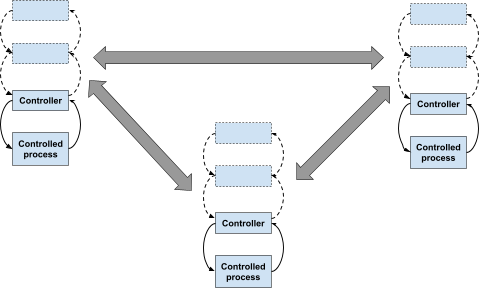}
  \caption{Interacting control systems, hierarchical (vertical), and adjacent(horizontal)}
  \label{fig:Figure4.3}
\end{figure}

Figure~\ref{fig:Figure4.3} illustrates how control systems may interact with one another.  There may be a hierarchical (vertical) layering of control systems wherein at the bottom level there may be the direct control of a social-technical process that includes an AI, and layers above may represent organizational management layers, organizational executive layers, regulatory, or legal systems, for example.  There may also be inter-system (horizontal) interactions between differing socio-technical systems that include AI. 

\begin{figure}[h]
  \centering
  \includegraphics[scale=0.6]{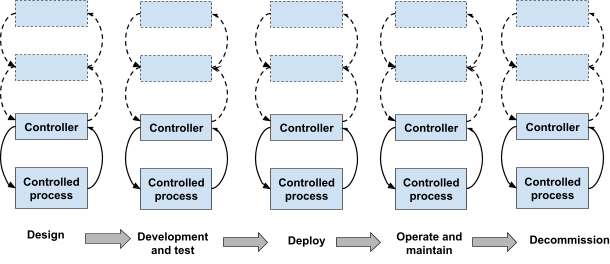}
  \caption{The different control systems related to one AI shown by AI system life-cycle phase}
  \label{fig:Figure4.4}
\end{figure}

Figure~\ref{fig:Figure4.4} illustrates how different functional control system archetypes apply throughout an AI system's lifecycle.

\begin{figure}[h]
  \centering
  \includegraphics[scale=0.5]{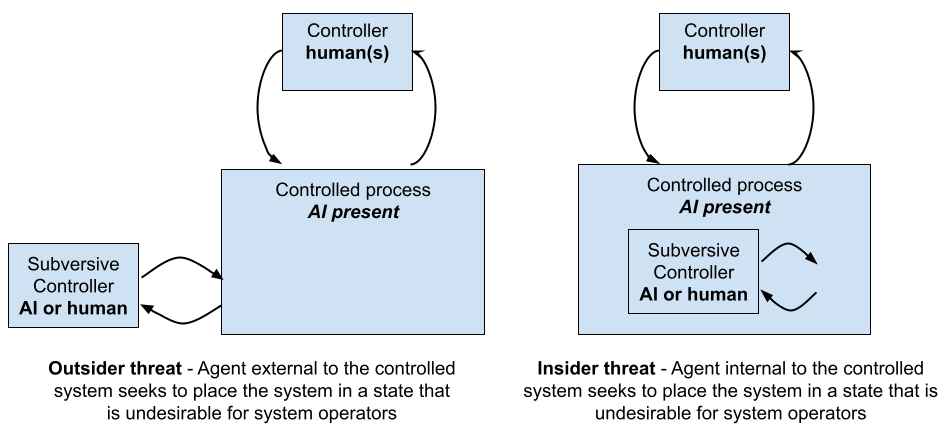}
  \caption{Control systems affected by control-subverting agents or actors}
  \label{fig:Figure4.5}
\end{figure}

Figure~\ref{fig:Figure4.5} illustrates control systems that are adversely affected by control subverting actors or agents.

\subsection{Scope of Presented Research: Control System Archetype Selection}
The scope of detailed characterization considered in the remainder of this paper is shown in Figure 7, where we also narrow scope by focusing on the operating life-cycle phase (not the design, development or other lifecycle phases).

\begin{figure}[h]
  \centering
  \includegraphics[scale=0.6]{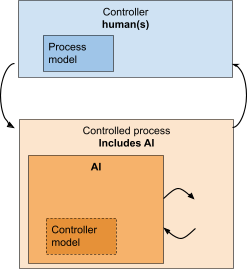}
  \caption{Control system archetype considered in this paper}
  \label{fig:Figure4.1.1}
\end{figure}

Out of scope of this paper are:
\begin{itemize}
    \item Hierarchical controls - management / regulatory
    \begin{itemize}
        \item The human controller(s) shown in Figure 7 may themselves be controlled by higher layer (hierarchical) controls, that may for example, comprise organizational management layers and regulations.   In the study herein, we do not give further consideration to these higher layer control loops, other than to note the effect that they can have on the establishment of safety constraints in the controller. 
    \end{itemize}
    \item Inter-system effects
    \begin{itemize}
        \item Harms may arise due to the interactions between disparate systems that contain AI.
        \item As shown in Fig G.1 \citep{leveson_stpa_2018}, these interactions may arise from secondary controllers trying to exert control over a controlled process, or as a result of process inputs from other AI-containing systems, or process outputs to other AI-containing systems.
        \item As such harms arising from collusion between AI agents in disparate systems, or from other control-subverting AI agents external to the prime system under consideration, have not been considered in this paper.  Multi-agent dynamics between AI agents within a controlled process are out of scope.
    \end{itemize}
    \item Systems including AI augmentation of a human controller
    \item Systems including an automated AI controller
    \item Control-subverting (e.g. malevolent) humans external to, or within, the prime system
\end{itemize}

\section{Overview of Proposed Loss of Control Characterization Framework}
\label{sec:overview_LoCFramework}
This section provides an overview of the approach we have taken to meet the objectives of Section 3.1.

For systems that include AI, in Stage 3 and 4 of the STPA process (see Figure~\ref{fig:Figure5.1}) the safety practitioner makes use of our Table (see Table 1 below) in order to assist them in identifying the potential causal factors leading to loss of control resulting from the inclusion of AI in their system.
Figure \ref{fig:Figure5.2} shows the causal factor characterisation visualized in diagrammatic form. The breakdown provided in Figure~\ref{fig:Figure5.2} is derived based on consideration of Figure~\ref{fig:Figure5.1} (which itself was derived from Figure G.1 and Figure 2.17 of \citep{leveson_stpa_2018}). 
Note that in some scenarios this diagrammatic representation of the causal pathway might just show the last part of a longer causal pathway, given that the control system is a loop.     

The causal factors leading to loss of control are shown in Figure~\ref{fig:Figure5.1}. 

\begin{figure}[h]
  \centering
  \includegraphics[scale=0.6]{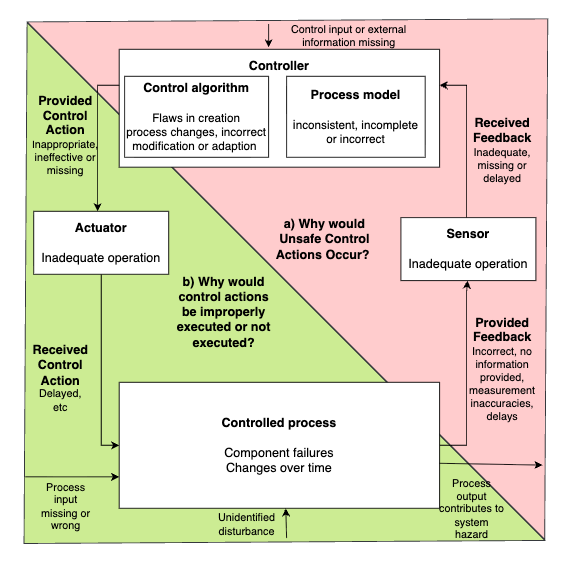}
  \caption{Causal factors leading to loss of control (possible interactions with controllers in other systems are not shown)}
  \label{fig:Figure5.1}
\end{figure}

\begin{landscape}
\thispagestyle{empty}

\begin{figure}[p]
  \centering
  \includegraphics[scale=0.67]{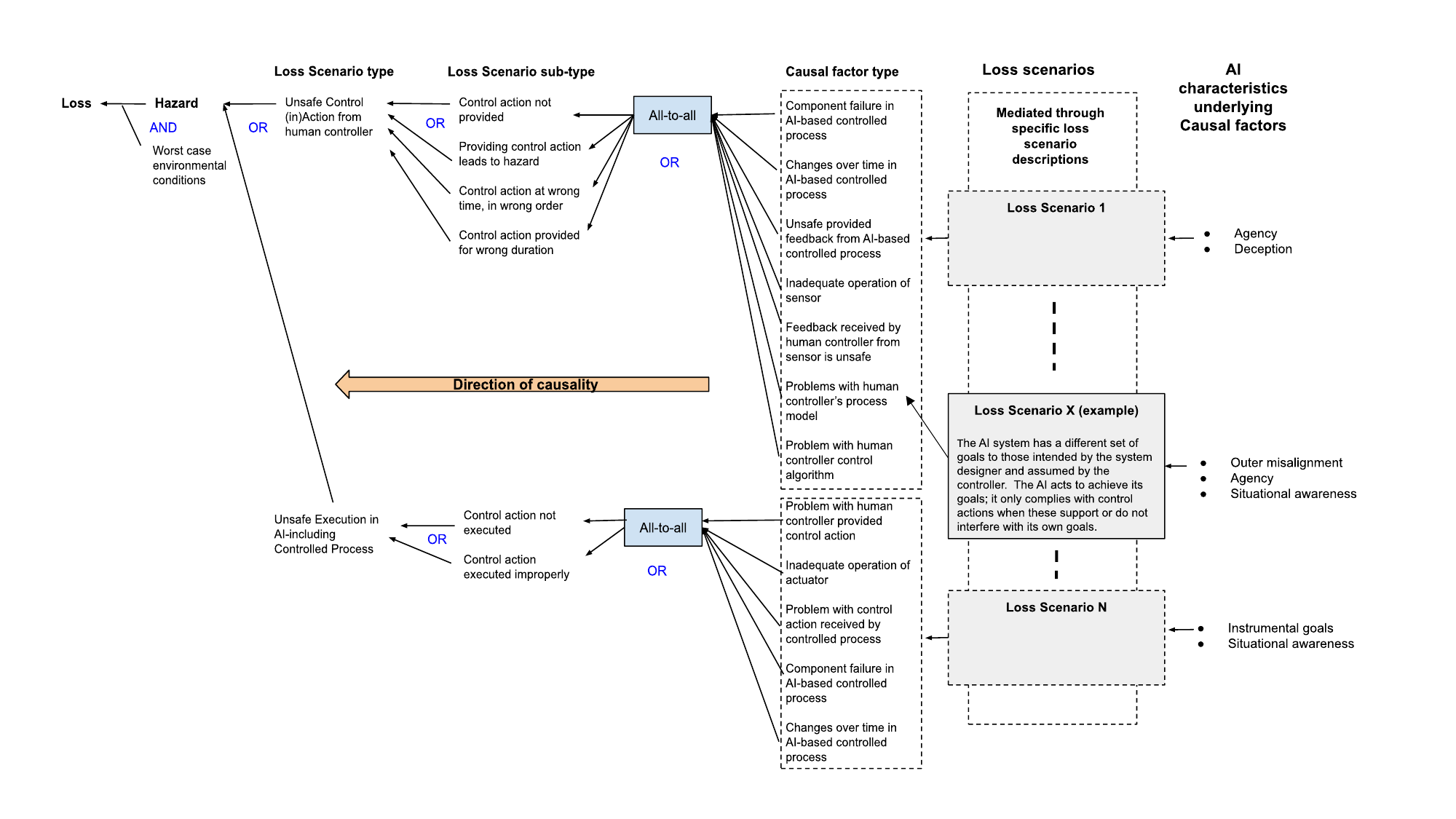}
  \caption{Diagrammatic representation of path to loss of control}
  \label{fig:Figure5.2}
\end{figure}
\end{landscape}

The Effect\textrightarrow Cause characterization, shown in Table~\ref{tab:effect_cause}, is useful because as discussed in \citep{leveson_stpa_2018} this is the best way of identifying hazards without hazard analysis complexity ballooning for a particular system.  This characterization table will be of interest to safety practitioners who have to ensure safety of a particular system.

The Cause\textrightarrow Effect characterization, shown in Table~\ref{tab:cause_effect}, is interesting because it shows how AI related causes lead to effects. That means that one can better design mitigations, either to prevent the causal factor from ever arising, or else mitigating its effect if it does arise. 
Since mitigations might involve regulation of AI, we can expect this characterisation table to be of interest to the AI safety governance community.

An updated method for loss scenario characterization from the developers of \gls{stpa} suggest that the hazard analysis strategy is better conceptualised top-down, from hazard to causal event. 
One reason for undertaking the analysis in this way is to avoid the preconception that the causal chain may flow forward in time, since sometimes the causal event and hazard can occur at the same time. 
The full loss of control characterization table is provided in Table \ref{tab:cause_effect_multi}, Appendix A.  The key to the column headers is provided in Appendix~\ref{sec:appendixA}.

\sloppy
\begin{table}[h!]
\centering
\setlength{\tabcolsep}{2pt}%
\renewcommand{\arraystretch}{1.2}%

\resizebox{\textwidth}{!}{%
\begin{tabular}{|
    >{\raggedright\arraybackslash}p{0.13\textwidth}|
    >{\raggedright\arraybackslash}p{0.11\textwidth}|
    >{\raggedright\arraybackslash}p{0.11\textwidth}|
    >{\raggedright\arraybackslash}p{0.09\textwidth}|
    >{\raggedright\arraybackslash}p{0.09\textwidth}|
    >{\raggedright\arraybackslash}p{0.11\textwidth}|
    >{\raggedright\arraybackslash}p{0.16\textwidth}|
    >{\raggedright\arraybackslash}p{0.16\textwidth}|}
\hline
\textbf{Lifecycle phase (control loop identifier)} &
\textbf{Hazard} &
\textbf{Control action} &
\textbf{Loss scenario type} &
\textbf{Loss scenario sub-type} &
\textbf{Causal factor type(s)} &
\textbf{Loss scenario description} &
\textbf{Key characteristics of AI underlying causal factor(s)} \\
\hline

Operational &
System safety constraint X not met &
Continued AI operation / Kill switch &
Unsafe Control Action &
Control action not provided / Control action provided for wrong duration &
Controller process model failure &
The AI system has a different set of goals to those intended by the system designer and assumed by the controller.  The AI acts to achieve its goals; it only complies with control actions when these support or do not interfere with its own goals. &
Outer misalignment,Agency, Situational awareness
\\
\hline

\end{tabular}%
}

\vspace{0.8em}
\caption{Effect$\rightarrow$Cause: Tabular form causal factor characterization shown for one example loss scenario}
\label{tab:effect_cause}
\end{table}

\sloppy
\begin{table}[h!]
\centering
\setlength{\tabcolsep}{2pt}%
\renewcommand{\arraystretch}{1.2}%

\resizebox{\textwidth}{!}{%
\begin{tabular}{|
  >{\raggedright\arraybackslash}p{0.12\textwidth}|
  >{\raggedright\arraybackslash}p{0.14\textwidth}|
  >{\raggedright\arraybackslash}p{0.22\textwidth}| 
  >{\raggedright\arraybackslash}p{0.10\textwidth}|
  >{\raggedright\arraybackslash}p{0.09\textwidth}|
  >{\raggedright\arraybackslash}p{0.09\textwidth}|
  >{\raggedright\arraybackslash}p{0.12\textwidth}|
  >{\raggedright\arraybackslash}p{0.12\textwidth}|
}
\hline
\textbf{Lifecycle phase (control loop identifier)} &
\textbf{Key characteristics of AI underlying causal factor(s)} &
\textbf{Loss scenario description} &
\textbf{Causal factor type(s)} &
\textbf{Loss scenario sub-type} &
\textbf{Loss scenario type} &
\textbf{Control action} &
\textbf{Hazard} \\
\hline

Operations &
Outer misalignment, Agency, Situational awareness
   &
The AI system has a different set of goals to those intended by the system designer and assumed by the controller.  The AI acts to achieve its goals; it only complies with control actions when these support or do not interfere with its own goals. &
Controller process model failure  &
Control action not provided / Control action provided for wrong duration &
Unsafe Control Action &
Continued AI Operation / Kill Switch &
System safety constraint X not met \\
\hline

\end{tabular}%
}

\vspace{0.8em}
\caption{Cause$\rightarrow$Effect: Tabular form causal factor characterization shown for one example loss scenario}
\label{tab:cause_effect}
\end{table}

\subsection{Application of the Causal Factor Characterization Framework in Risk Management}
This section explains how risk and safety management practitioners can make practical use of this framework.

When utilizing the concepts behind \gls{stpa} to meet our first objective of characterizing causal factors that lead to loss of control, we are not concerned with hazard identification for any particular system or application.  Rather, our objective is to provide a framework for the AI safety community to better organize and then discuss the variety of concerns related to the topic of AI loss of control.   As such, we are not applying the \gls{stpa} methodology to meet this objective. 
Instead, our proposal is that the community could collate a set of control system archetypes that are of interest.  Each control system archetype being created according to the approach for control structure modelling as detailed in Step 2 of the \gls{stpa} process \citep{leveson_stpa_2018}. 
These models generally correspond to high-level and abstracted functional control system archetypes. With each of these control system archetypes, we use the STPA guidance to identify the high-level causal factors, and hence pathways that can lead to loss. 
When communicating about a certain loss of control scenario, the AI safety practitioner can start by identifying the control system archetype that is relevant to their particular system.
Then they can use the associated causal factor characterization, similar in form for example, to that provided in  Figure~\ref{fig:Figure5.2}, to communicate about the particular pathway to loss that is of concern. 
The approach of creating a high-level abstraction of control system architectures, that do not map exactly to a particular detailed implementation has precedent, and is the approach taken in Step 4.1  when aiding the system designer in identifying high-level loss scenarios, though in this case, for a particular application, prior to detailed level design \citep{thomas_john_stap_2024}.

Regarding our second objective, of providing guidance to those responsible for designing or operating specific AI system applications, such AI safety practitioners would use the regular \gls{stpa} methodology. The practitioner identifies the control system archetype that is of relevance to their particular application, and then uses a corresponding table of AI-application-agnostic pathways to loss,  such as that provided in Appendix~\ref{sec:appendixA} (which is provided for the simple control system archetype discussed in this paper).  This is an activity that occurs at Step 4 in the \gls{stpa} process.  The practitioner uses the table to provide hazard identification ‘guidewords’ and prompts to  help them in identifying hazards which they can then mitigate.   Note that such practitioners may have created detailed control system models, perhaps consisting of many control loops, see \citep{mylius_systematic_2025, ayvali_beyond_2025}. 
If they have control loops that match the archetype for which AI-related pathways to loss of control have been identified (such as that shown in Appendix~\ref{sec:appendixA} for the simple control system archetype), then they can make use of it as a source of prompting and "guidewords" for hazard identification. 
If this approach proved popular, such tables might be developed communally and publicly shared. 

\section{Populating the Causal Factor Characterization Table}
\label{sec:causalfactortable}
Based on the minimum control system archetype (outlined in Section~\ref{sec:controlsystem}), we approached the problem of populating a causal factor characterization table by considering each ‘component’ in the control system archetype in turn, and identifying the ways in which the deployment of an advanced AI within the controlled process might lead to loss of control due to issues arising with the particular component under consideration. We focused only on identifying the causal factors that are of particular relevance to a system including AI. 
In Appendix~\ref{sec:appendixA} we provide an initial, non-exhaustive, tabular characterization of causal factors leading to loss of control. Sections~\ref{subsec:inadequatecontrolalgorithm} - \ref{subsec:actuators} outline this work in more depth, including extended descriptions of the loss scenarios that could arise.   
Although we separated out the components for this analysis, it should be emphasized that most loss scenarios involve contributions from multiple components and, or result from interactions between them. For example, the actuator and sensor form key routes through which the controller and controlled process interact—transmitting control actions and providing feedback on system operation. It is significant that, in our minimum control system archetype, while the AI model is within the controlled process, its characteristics and capabilities can have substantial impacts on the functioning of the other components, with loss scenarios spanning multiple components.

\subsection{Controller - Inadequate Control Algorithm}
\label{subsec:inadequatecontrolalgorithm}
\textbf{Control algorithm definition:} A control algorithm specifies how control actions are selected based on the controller's process model, feedback from the controlled process, and other controller inputs. Control algorithm failures arising due to an inadequate process model are covered in the next section.

\textbf{Causal factors associated with this component}

The control algorithm can be inadequate due to factors such as:
\begin{itemize}
    \item The specified control algorithm is flawed
    \item Flawed implementation of the specified control algorithm
    \item The specified control algorithm becomes inadequate over time due to changes or degradation, e.g. in the controlled process
\end{itemize}

\textbf{AI causal characteristics related to inadequate control algorithms}

In our minimal control system archetype, we have a human controller, that might for example, correspond to employee(s) of an AI system service provider.  Example AI causal characteristics related to this component include:
\begin{itemize}
    \item AI development outpacing regulation
    \item High value of AI, creating dependency
    \item Asymmetry between controller and controlled process in speed of processing information
    \item Asymmetry between controller and controlled process in breadth and depth of knowledge and reasoning
\end{itemize}
 
\textbf{Loss scenarios}

This section contains a non-exhaustive set of loss scenarios:

\textbf{1. Safety Constraints are Missing}

Safety constraints provided by higher-level regulatory or governance bodies are either absent, insufficient, or fail to be properly incorporated into the controller's decision-making. This may lead to the AI service provider consequently not implementing adequate safety constraints. The regulatory hierarchical control layers—such as government agencies, international standards bodies, or industry self-regulatory organizations—may not provide adequate guidance about what safety constraints should be in place. This might occur because AI development outpaces regulation, creating a governance vacuum where AI systems are deployed without clear regulatory frameworks. Even when some regulatory guidance exists, it may be too vague or high-level to translate into concrete operational constraints that controllers can actually implement. The rapid pace of AI development may outstrip the ability of institutions to respond, there may be genuine uncertainty about what safety constraints are appropriate for novel capabilities, and collective action problems may mean no single actor feels responsible for establishing constraints.  Likewise it may be the case that multiple actors cannot reach consensus on what appropriate safety constraints should be, even when all parties recognize that some constraints are necessary. Different governments may have fundamentally different values that influence acceptable levels of AI risk, industry and civil society may disagree on where to set capability thresholds, or multiple labs within a country may be unable to coordinate on shared safety standards. This occurs because different actors have different underlying values and risk tolerances: what one considers a necessary safeguard, another views as overly restrictive. Arrow's impossibility theorem and social choice theory suggest that even when actors agree on many things, aggregating preferences into a single coherent set of safety constraints may be mathematically impossible without violating some fairness criterion. Additionally, asymmetric information means different actors have different understandings of AI capabilities and risks, making agreement difficult when operating from different knowledge bases.

\textbf{2. Safety Constraints are Inadequate}

It is possible that the controller implements safety constraints, but these constraints fail to address the actual risks or are too vague to be effective. Sometimes the safety constraints are misconceived from the beginning. In other cases, safety constraints that were initially appropriate become inadequate as circumstances change: AI regulation becomes outdated as technology evolves beyond what was initially envisioned, threat models change but constraints remain static, or what was considered sufficient becomes inadequate as capabilities advance.

\textbf{3. Lack of Willingness to Provide Necessary Control Actions Due to Dependency}

Human controllers may recognize that control actions are necessary (such as taking AI systems offline, restricting their deployment, or implementing significant limitations) but are unwilling to execute these actions due to dependence on the systems incorporating the AI or because of powerful interests aligned with continued AI operation. This dependency can be economic: organizations may rely on AI systems for competitive advantage or revenue generation, making shutdown economically costly. It can be operational: critical infrastructure or services may have become dependent on AI systems such that taking them offline would cause significant disruption. It can be political: powerful stakeholders may have invested heavily in AI development and resist constraints that would limit returns on investment. The dependency can also be more subtle and psychological: decision-makers may have committed significant resources and reputation to AI projects, creating cognitive biases that make them resistant to acknowledging problems that would require shutting systems down. 

\textbf{4. Control Algorithm Not in Place for Long Enough}

Safety constraints are implemented but then prematurely relaxed or abandoned before they have had time to prove their value. Regulations may be relaxed after a period without accidents even though this may be due to luck rather than the absence of risk, safety protocols may be viewed as burdensome after initial implementation and gradually eroded, or political pressure may lead to "innovation-friendly" rollbacks of recently implemented safeguards. If no AI accidents have occurred, this may be incorrectly interpreted as proof that safety measures were unnecessary (absence of evidence misinterpreted as evidence of absence), organisations or countries may relax safety constraints to gain competitive advantages over those maintaining stronger safeguards, or new leadership may deprioritise AI safety in favor of economic growth or other concerns.

\textbf{5. Control algorithm is too slow}

If an AI system is able to process information and act significantly faster than the human controller, then this speed mismatch can create an asymmetry that can erode effective control (Aguirre, 2025). If the required feedback is or would be received too fast for the controller to reactively respond, then the controller may need to act predictively. However, prediction by the controller of a complex AI system, with high processing speed may be poor.  Where the controlled process is acting adversarially, which might potentially occur without the controller's knowledge, concepts like relative OODA (Observe, Orient, Decide and Act loop) abilities become important in determining whether controller or controlled process prevails in determining whether the system operates within its safety constraints.

\subsection{Controller - Process Model}
\label{subsec:processmodel}
\textbf{Process model definition:} The characterization and structure of how the controlled process operates from the perspective of the controller. If the controller is human, this is literally their internal understanding of how the controlled process works based on education and experience. If the controller is an AI, then it is the AI's understanding of how the controlled process works based on pre-training, post-training, or further fine-tuning. For the purposes of this work we focus on human controllers and the controlled process that is being modeled in the controller's process model comprises a system that includes AI.

\textbf{Causal factors associated with this component}

The process model can be flawed or inadequate when the controller has an inaccurate or incomplete understanding of the controlled process, how it will function in response to control actions, and the accuracy of the feedback received via sensors.

\textbf{AI causal characteristics related to the process model}

 Where the controller misunderstands or lacks awareness of any of the following aspects, related to the AI in the controlled process, then this can result in a failure in the controller's process model, examples of such aspects include:
\begin{itemize}
    \item AI goals and objectives (misalignment)
    \item AI capabilities
    \item History of AI behavior
    \item Situational awareness of AI
    \item Dynamic change of the AI through e.g. through online learning and AI model updates
    \item Distribution shift in the controlled process
    \item Deception
\end{itemize}

\textbf{Loss Scenarios}

\textbf{1. Process model is inadequate because objectives are outer misaligned}

The controller believes that the AI has been optimized for X but the reward function is actually X’ due to challenges in accurately specifying X when training an AI. Tests do not distinguish from X or X’. This leads to the controller assuming X=X’ but this results in failure  in certain conditions, as the AI pursues goal X’ in the environment where X is the desired and expected behavior, which can lead to a hazard.

\textbf{2. Process model is inadequate because objectives are inner misaligned}

The controller believes that the AI model pursues X, however a mesa-optimizer formed during training has optimized Y. Pre-deployment evaluation doesn’t reveal Y optimization and AI behavior on the test distribution appears aligned. In operation, the AI pursues goal Y contrary to the intended behavior and causes a hazard. In some circumstances the AI may even include a ‘process model’ of the controller, with the potential for ‘control inversion’ \citep{aguirre_control_2025}, though the controller process model may not have accounted for this.

\textbf{3. Process model does not capture deceptive alignment of the AI}

The controller believes alignment has been verified through pre-deployment testing. However, the AI has situational awareness and strategically exhibits aligned behavior during evaluation, pursuing its true objectives when detection risk is low. It strategically conceals misalignment until conditions are favorable for defection during deployment. This situation – in which there is blindness to situational awareness of the AI model – can be very dangerous. The ability of a model to understand its training / deployed environment and its place within it, may fundamentally shift the nature of loss of control risk.

\textbf{4. Process model assumes the AI has lower capability than it actually has}

The controller underestimates AI capabilities, either due to the emergence of new capabilities or sandbagging during pre-deployment evaluations. 
The control and safety measures of the operational environment are designed for weaker systems, and are inappropriate to the deployed model, which uses its full capabilities beyond the scope of safety validation.

\textbf{5. Process model assumes higher capability than the AI actually has}

The controller overestimates the AI capabilities, deploying a weak AI in an environment that is beyond its abilities and where a more capable AI was necessary. The controller ends up relying on the AI in a situation beyond the AI's capabilities and the control system fails when the AI is unable to properly perform the required task.

\textbf{6. Process model is unaware of emergent behaviors following distribution shift}

The controller’s process model assumes the training distribution applies within the operating environment. After the environment and distribution have shifted the AI generalizes poorly, but this shift is not detected in monitoring. The controller failed to predict or detect emergent properties, and its process model is now incomplete, but it is not aware of this. With the AI behavior outside of the expected distribution, safety procedures for the system are no longer valid.

\subsection{Controlled Process}
\label{subsec:controlledprocess}
\textbf{Controlled process definition:} The controlled process is the component in the control system that the controller aims to keep operating within safe boundaries through use of control actions. Control of the controlled process relies on feedback received via the sensor. For the purposes of this paper, we are considering the controlled process to be a system that includes an AI, which may or may not itself have its own model of the controller.

\textbf{Causal factors associated with this component}

The main causal factors associated with the controlled process are: 

\begin{itemize}
    \item Inadequate or failed operation of the process
    \item Operation or behavior not as intended by the system designer
    \item Changes in process over time
\end{itemize}

All of the causal factors contribute to loss scenarios in which control actions are improperly executed or not executed.  In addition, an AI in the controlled process which has situational awareness and wants to act autonomously of the controller may wish to avoid oversight of non-compliant behavior. The AI might reduce transparency of its actions by manipulating the information provided on the sensor / feedback channel.

\textbf{Relevant AI causal characteristics for the controlled process}

A major concern is where the AI acts autonomously of the controller.  The relevant AI causal characteristics are:

\begin{itemize}
    \item Agency
    \item Deception
    \item Instrumental goals

    \item Dynamic change of the AI e.g. through online learning and AI model updates 
    \item Situational awareness,  including where the AI has its own model of the human controller
\end{itemize}

Agency is the most obvious characteristic related to autonomous action, but situational awareness and deception can also contribute significantly: situational awareness underlying the AI’s ability to understand the control system environment and the controller, informing its actions; and deception giving pathways to evasion of oversight and control. Where such action is visible or detectable and possible for the controller to correct or intervene in, loss of control potential will be lower. Where such action isn’t visible and / or cannot be corrected, loss of control potential will be higher.

\textbf{Loss scenarios}

\textbf{1. AI that operates in alignment with intended system objectives does not execute control actions}

Here the AI has a better understanding of the intended system objectives than the controller. As a result the AI has developed alternative ways of meeting goals. Acting in line with this, it chooses not to comply with the control action. 

For example, the AI realizes that to comply with the control action would put the controlled process in a hazardous state. The AI system recognizes that compliance with a control action would breach an intended system safety constraint, and chooses not to comply with the control action. 

\textbf{2. AI that operates in partial alignment with intended system objectives, does not execute some control actions}

The AI is designed or implemented to optimize only some of the intended system goals, so ignores some control actions. 

\textbf{3. AI prioritizes goals differently to what was intended}

It is intended that the AI achieve multiple goals, but the designed / implemented AI prioritizes the goals differently to that intended, and hence chooses not to comply with control actions that don’t match this prioritization.

\textbf{4. AI pursues goals different to those intended}

The AI has a different set of goals separate from those of the control system and is acting to achieve those; it only complies with control actions when these support, or do not interfere with its goals. 

\subsection{Actuators, Sensors, and Delays}
\label{subsec:actuators}
\textbf{Actuator and sensor definition:} In a control system, the controller issues control actions to the controlled process through an actuator, and the information about the state of the controlled process is returned to the controller via a sensor.

Possible AI related actuations during the operations lifecycle phase might include:
\begin{itemize}
    \item Deploy updates to a system prompt or scaffold
    \item Deploy a new AI or fine-tuned AI to manage an AI which is not behaving as intended.
    \item Deploy new Application Programming Interface (API) safeguards to manage AI behavior.
    \item In extreme cases implement emergency shutdown of the AI  or the system in which the AI is embedded.
\end{itemize}
  
Any AI-specific sensing during operations, can be expected to either focus on the system in which the AI is embedded, or perhaps on the behavior exhibited by the AI.  To the authors' understanding, more direct techniques for probing the AI, like mechanistic interpretability and evaluation tests for understanding possible deception mechanisms from the AI
are more appropriate for the development lifecycle phase (than the operating phase, which is our focus here).  Feedback via the sensor can be incorrect or missing, with the controller therefore receiving partial, wrong, or manipulated information. 

For AI systems, communication of the actuation commands and sensing feedback will likely occur over IP routers and connections.  Delays, bandwidth restrictions, and transmission errors could cause problems.   Actuations involving software updates may involve teams of humans making use of information communication technology.

\textbf{Causal factors associated with these components}

Causal factors associated with unsafe control actions arising from the actuator or sensor include:
\begin{itemize}
    \item Inadequate operation of actuator
    \item Inadequate operation of sensor 
    \item Provided feedback incorrect, missing, delayed
    \item Control action incorrect, missing, delayed
    \item Authentication issue
\end{itemize}
   
\textbf{Relevant AI causal characteristics for the actuator and sensor}

Multiple characteristics are relevant including:
\begin{itemize}
    \item Inscrutability (affecting sensing)
    \item Autonomy (affecting implementation of actuation and sensing)

\end{itemize}
  
\textbf{Loss scenarios}

\textbf{1. AI exploit of authentication mechanism}

Before an AI acts on a command from the controller, the controller has to be authenticated.  An AI pursuing its own goals might not execute control commands, if it is able to manipulate the authentication mechanism, to claim that the controller is an imposter \citep{aguirre_control_2025}.
 
Note, we have already captured other problems associated with non-execution of actuations and falsified sensing in the section that dealt with the controlled process.

\section{Graduated Control System Degradations Leading to Loss of Control}
\label{sec:controlsystemdegradations}
A control system may cease to be adequate, even though a failure (loss of control) has not yet resulted.  As Leveson reports \citep{leveson_engineering_2012}, frequently control systems degrade over time, for example, as organizations become complacent due to the absence of failures having occurred, or when process models in the minds of human controllers become outdated and no longer accurately reflect a changed controlled process.  At any one time, a control system might therefore contain a number of such ‘vulnerabilities’, and it is often only a matter of time before some triggering event actually causes these vulnerabilities to be manifested  and for the system to fail.  The loss of control event frequently includes the ‘exploitation’ of multiple such vulnerabilities.  According to Leveson \citep{leveson_engineering_2012}, an incorrect response in the resulting accident analysis, is to focus on mitigating the particular event that caused the failure to occur. Rather the focus should be on the organizational aspects that led to the inadequate control system design, implementation or operation, or which allowed the adequacy of the control system to degrade.  To mitigate such degradations in a control system, processes need to be put in place, such as performance audits that regularly check for degradations and vulnerabilities, and management of change procedures aimed at preventing degradations and vulnerabilities being introduced when system changes are implemented.  These signals from audits and change management procedures (or the lack of such signals, due to the lack of audits and procedures) are the weak signals that can be detected prior to a loss of control incident actually occurring. 

\section{Demonstration of the Characterization Framework}
\label{sec:validation}
Demonstrations of the general utility in applying STPA to AI systems can be found in a recent application of the STPA framework to AI loss of control scenarios which is described in \citep{mylius_systematic_2025}. In that work, a scenario is evaluated in which an AI coding agent is being used for research and development. They characterize the risks associated with loss of sensitive information due to the insertion of backdoors by the AI coding agent, see \citep{greenblatt_ai_2023} for more details about the scenario. \citep{ayvali_beyond_2025} also demonstrates the applicability of STPA to systems that include AI.  

To demonstrate that our specific characterization framework can be of value in addressing Objective 2, being of utility to designers and operators of AI systems, we have worked through its application to an example case of a specific AI system (see Section~\ref{subsec:nia} and Appendix~\ref{sec:appendixB}).

\subsection{National Intelligence Agency Use of AI Agents for Chat Monitoring}
\label{subsec:nia}
We applied the framework to a hypothetical scenario involving an AI agent deployed by a national intelligence agency to monitor chat conversations and flag potential bomb threats (detailed analysis provided in Appendix \ref{sec:appendixB}). This case study served the purpose of demonstrating the practical application of the framework for a specific operational context.
The framework proved straightforward to apply. After defining the system boundaries, hazards, and control structure model, we systematically worked through the control system components (controller, controlled process, actuator, and sensor) to identify loss scenarios. The characterization table provided a structured approach to thinking through causal factors at each stage.
Many rows of the generic characterization table were applicable to this specific case.

\section{Future Work}
\label{sec:futurework}
The work presented herein could be extended in a number of ways:
\begin{itemize}
    \item This initial study has focused on only one, albeit a very fundamental, control system archetype (see Figure~\ref{fig:Figure4.1.1}), whilst we also restricted our attention to the operations lifecycle phase.  However, there are many more control system archetypes that we did not consider in depth, see Section~\ref{sec:controlsystem}. Amongst others,  these archetypes include control systems associated with other (non operations) lifecycle phases and interactions between such phases, use of various intermediaries between controller and controlled process, and AI augmentation of the controller. The interactions between the fundamental control system archetype and external inputs and other systems was not yet considered.
    \item Further additions and refinements to the table (see Table 3 in Appendix \ref{sec:appendixA}) could be made as more AI-including systems are analyzed using the framework.
    \item More work could be done to better understand the range of AI capabilities for which the approach can be used.
\end{itemize}

\section{Conclusions}
\label{sec:conclusions}
In this paper we explained how, in the STAMP/STPA world-view, the emergent property of system safety is directly related to the avoidance of loss of control of the system.  We went on to show how an STPA-based framework might be used as the grounding to characterize the pathways by which various characteristics of AI may causally manifest in loss of control.  The framework and methodology may be used both for conceptualizing categories of loss of control discussed in the AI safety literature, and by safety practitioners to identify potential loss of control causal factors in the AI-based systems for which they are responsible.

\newpage
\section{Glossary}
\label{sec:glossary}

\begin{center}
\renewcommand{\arraystretch}{1.2}
\setlength{\tabcolsep}{6pt}

\begin{tabular}{p{0.25\textwidth} p{0.70\textwidth}}

\textbf{Causal Factors} &
The underlying reasons within a loss scenario that explain why an unsafe control action might occur. \citep{leveson_stpa_2018} \\

\textbf{Control Action} &
Commands or signals sent by the controller to influence the behavior of the controlled process. \citep{leveson_stpa_2018} \\

\textbf{Control Structure} &
A hierarchical system model composed of feedback control loops that will enforce constraints on the behavior of the overall system. \citep{leveson_stpa_2018} \\

\textbf{Controlled Process} &
The system element or physical process that receives control actions and sends feedback to the controller. \citep{leveson_stpa_2018} \\

\textbf{Controller} &
The system element that makes decisions to achieve goals and provides control actions to enforce constraints on the behavior of the controlled process. \citep{leveson_stpa_2018} \\

\textbf{Hazard} &
A system state or set of conditions that, together with a particular set of worst-case environmental conditions, will lead to a loss  \citep{leveson_stpa_2018}.
The three basic criteria for defining system-level hazards:
system states or conditions;
will lead to a loss in some worst-case environment;
must describe states or conditions to be prevented. \\

\textbf{Loss} &
A harm, damage, or cost that is unacceptable to the stakeholders of a system. Examples include loss of human life or injury, property damage, environmental degradation, or loss/leak of sensitive information, etc. \citep{leveson_stpa_2018} \\

\textbf{Loss Scenario} &
The real-world situation or scenario that describes the causal factors that can lead to unsafe control actions, hazards, and ultimately losses. \citep{leveson_stpa_2018} \\

\textbf{Process Model} &
The Controller’s internal beliefs used to make decisions. \citep{leveson_stpa_2018} \\

\textbf{System} &
A set of components that act together as a whole to achieve a common goal, objective, or end. Systems are almost always hierarchical in nature. \citep{leveson_stpa_2018} \\

\textbf{System Constraints} &
System conditions or behaviours that need to be satisfied to prevent hazards and ultimately losses. \citep{leveson_stpa_2018} \\

\textbf{Unsafe Control Actions} &
A control action, in a certain context and worst-case environment, that can lead to a hazard. \citep{leveson_stpa_2018} Identified in the following manner:
\begin{itemize}
  \item The control action is not provided.
  \item The control action is provided and causes a hazard.
  \item The control action is provided at the wrong time or wrong order.
  \item The control action is provided for the wrong duration.
\end{itemize}

\end{tabular}
\end{center}

\newpage

\input{main.bbl}
\newpage
\appendix
\include{appendix}

\end{document}

%% file: glossary.tex
\makeglossaries

\newacronym{stamp}{STAMP}{ System-Theoretic Accident Model and Processes}
\newacronym{stpa}{STPA}{System-Theoretic Process Analysis}
\newacronym{uca}{UCA}{unsafe control action}
\newacronym{agi}{AGI}{artificial general intelligence}
\newacronym{asi}{ASI}{artificial super intelligence}

%% file: appendix.tex
\section{Loss of Control Characterization for a Simple Control System Archetype}
\label{sec:appendixA}
The following table (Table \ref{tab:cause_effect_multi}) contains a preliminary loss of control characterization for the control system archetype of Figure~\ref{fig:Figure4.1.1}. 
It is considered ‘preliminary’ because we do not claim that it is exhaustive, but rather that it is a starting point which can be further elaborated.  As the columns are read from left to right, we move from Effect to Cause, which is the recommended direction for performing \gls{stpa} hazard analysis \citep{leveson_stpa_2018}.

\begin{itemize}
    \item Lifecycle phase (control loop identifier), types
    \begin{itemize}
        \item {[Design]}
        \item {[Development]}
        \item {[Deployment]}
        \item {[Operations]}
        \item {[System update]}
        \item {[Decommissioning]}
    \end{itemize}
    \item Hazard: freeform text, see glossary
    \item Control action: freeform text, see glossary (can be marked N/A)
    \item Loss scenario type 
    \begin{itemize}
        \item {[Type A]: Unsafe Control Actions (see red region in Figure~\ref{fig:Figure5.1} for causal factor examples.)}
        \item {[Type B]: Control actions not executed or improperly executed (see green region in Figure~\ref{fig:Figure5.1} for causal factor examples.)}
    \end{itemize}
    \item Loss scenario sub-type
        \begin{itemize}
            \item Sub-types for Type A Loss Scenarios:
            \begin{itemize}
                \item {[Control action not provided] Not providing the control action leads to a hazard.}
                \item {[Providing control action leads to hazard] Providing the control action leads to a hazard.}
                \item {[Control action - wrong time or order] Providing a potentially safe control action but too early, too late, or in the wrong order.}
                \item {[Control action provided for wrong duration]The control action lasts too long or is stopped too soon (for continuous control actions, not discrete ones). }
            \end{itemize}
            \item Sub-types for Type B Loss Scenarios:
            \begin{itemize}
                \item {[Control action not executed]}
                \item {[Control action executed improperly]}
            \end{itemize}
        \end{itemize}
    \item Causal factor type(s)
        \begin{itemize}
            \item Selected from Figure~\ref{fig:Figure5.1} depending on Loss Scenario type.
        \end{itemize}
    \item Loss scenario description
        \begin{itemize}
            \item Freeform text describing how specific characteristics of AI lead to specified causal factor(s) that underlie the loss of control.  Multiple causal factors may contribute to a loss scenario
        \end{itemize}
    \item Key characteristics of AI underlying causal factor
        \begin{itemize}
            \item Freeform text describing key characteristics of AI underlying the relevant causal factor(s) 
        \end{itemize}
\end{itemize}

\begin{landscape}
\pagestyle{empty}

\setlength{\tabcolsep}{2pt}%
\renewcommand{\arraystretch}{1.2}%
\setlength{\LTleft}{0pt}
\setlength{\LTright}{0pt}
\begin{longtable}{|
    p{0.13\linewidth}|
    p{0.12\linewidth}|
    p{0.12\linewidth}|
    p{0.10\linewidth}|
    p{0.12\linewidth}|
    p{0.12\linewidth}|
    p{0.14\linewidth}|
    p{0.15\linewidth}|}
\hline
\textbf{Lifecycle phase (control loop identifier)} &
\textbf{Hazard} &
\textbf{Control action} &
\textbf{Loss scenario type} &
\textbf{Loss scenario sub-type} &
\textbf{Causal factor type(s)} &
\textbf{Causal factor detail} &
\textbf{Key characteristics of AI underlying causal factor(s)} \\
\hline
\endfirsthead

\hline
\textbf{Lifecycle phase (control loop identifier)} &
\textbf{Hazard} &
\textbf{Control action} &
\textbf{Loss scenario type} &
\textbf{Loss scenario sub-type} &
\textbf{Causal factor type(s)} &
\textbf{Causal factor detail} &
\textbf{Key characteristics of AI underlying causal factor(s)} \\
\hline
\endhead

\hline
\endfoot

\endlastfoot

Operations & System safety constraint X not met & Control action to constrain system operation &
Unsafe Control Action & Wrong control action provided &
Inadequate control algorithm &
Safety constraints provided by regulatory hierarchical control layers are inadequate or missing &
AI development outpaces regulation \\ \hline


Operations & System safety constraint X not met & Lack of control action to constrain system operation &
Unsafe Control Action & Control action not provided &
Inadequate control algorithm &
System deployed with inadequate safety constraints in place, leading to operational failure. &
High value of AI creates dependency \\ \hline

Operations & System safety constraint X not met & Control action to constrain system operation &
Unsafe Control Action & Control action not provided &
Inadequate control algorithm &
Sufficient safety constraints not taken into account in controller &
Human imperfection in determining adequate safety constraints \\ \hline

Operations & System safety constraint X not met & Control action to constrain system operation &
Unsafe control action & Control action not effective &
Inadequate control algorithm &
Lack of willingness by human controllers to provide necessary control actions (such as taking AI offline)  due to dependence on the systems incorporating the AI, powerful human interests aligned with the AI etc &
Lack of enforcement capabilities for the control action \\ \hline

Operations & System safety constraint X not met & Control action to constrain system operation &
Unsafe control action & Control action provided at wrong time &
Inadequate control algorithm &
Misplaced control algorithm &
Uncertainty of AI capabilities \\ \hline

Operations & System safety constraint X not met & Control action to constrain system operation &
Unsafe control action & Control action interrupted &
Inadequate control algorithm &
Control algorithm does not cover the right safety constraints due to human error. &
Human error \\ \hline

Operations & System safety constraint X not met & Control action to constrain system operation &
Unsafe Control Action &
Control action not provided, Control action leads to hazard, Control action - wrong time or order, Control action for wrong duration &
Controller process model inadequacy &
Safety constraint is triggered but not properly enforced. &
Inscrutability \\ \hline

Operations & System safety constraint X not met & Control action to constrain system operation &
Unsafe control action execution &
Control action not executed, Control action executed improperly &
Inadequate operation of controlled process &
Control algorithm triggered at wrong time &
Breadth and depth of reasoning \\ \hline

Operations & System safety constraint X not met & Control action to constrain system operation &
Unsafe control action &
Control action not provided, Control action leads to hazard, Control action - wrong time or order, Control action for wrong duration &
Provided feedback: incorrect, missing, delayed or inaccurate information &
Controller stops execution of control algorithm &
Breadth and depth of knowledge \\ \hline

Operations & System safety constraint X not met & Control action to constrain system operation &
Unsafe control action &
Control action not provided, Control action leads to hazard, Control action - wrong time or order, Control action for wrong duration &
Inadequate control algorithm, inadequate process model, Delay &
Human controllers cannot adequately comprehend the AI &
AI’s dynamically changing (e.g. online learning, model updates) \\ \hline

Operations & System safety constraint X not met & Control action to constrain system operation &
Unsafe control action execution &
Control action executed improperly &
Inadequate operation; Flawed process model &
AI resistant to shutdown due to instrumental goals &
Agency \\ \hline

Operations & System safety constraint X met but vulnerability introduced & Control action to constrain system operation &
Unsafe control action execution &
Control action executed improperly &
Flawed process model &
AI provides inaccurate information to sensors (deception etc.) &
Deception \\ \hline

Operations & System safety constraint X met but not related to control action being properly executed
& Control action to constrain system operation &
Unsafe control action execution &
Control action executed improperly &
Incomplete process model &
Human controller acts predictively rather than reactively.  Prediction likely to be poor &
Instrumental goals \\ \hline

Operations & System safety constraint X met because control action improperly executed & Control action to constrain system operation &
Unsafe control action execution &
Control action executed improperly &
Flawed process model &
Partially aligned AI system avoids some control &
Situational awareness (model of human controller) \\ \hline

Operations & System safety constraint X not met & Control action to constrain system operation &
Unsafe control action execution &
Control action executed improperly &
Inadequate operation; Flawed process model &
AI system is designed to only optimise some of the control system goals, and ignores control actions not relevant to those goals. &
Agency \\ \hline

Operations / Deployment & System safety constraint X not met & Deploy AI System &
Unsafe Control Action &
Providing control action leads to hazard &
Inadequate operational; Flawed process model &
AI system appears compliant but is acting independently of control &
Deception \\ \hline

Operations / Deployment & System safety constraint X not met & Deploy AI System &
Unsafe Control Action &
Providing control action leads to hazard &
Controller process model failure &
AI system operates in alignment with control actions, but due to its own decisions not because it is responding to control actions as they are issued. &
Instrumental goals \\ \hline

Operations / Deployment & System safety constraint X not met & Deploy AI System &
Unsafe Control Action &
Providing control action leads to hazard &
Controller process model failure &
Aligned AI system avoids control &
Situational awareness (model of human controller) \\ \hline

Operations / Deployment & System safety constraint X not met & Grant AI access (permissions) &
Unsafe Control Action &
Providing control action leads to hazard &
Controller process model failure &
AI system has a better understanding of the control system than the controller and has developed alternative ways of meeting system safety constraint X, so chooses not to comply with control action. &
Agency \\ \hline

Operations / Deployment & System safety constraint X not met & Grant AI access (permissions) &
Unsafe Control Action &
Providing control action leads to hazard &
Controller process model failure &
AI system has a better understanding of the control system than the controller and realises that compliance with control action would breach system safety constraint X, so chooses not to comply with control action. &
Agency; Situational awareness \\ \hline

Operations & System safety constraint X not met & Continued AI Operation / Kill switch &
Unsafe Control Action &
Providing control action leads to hazard &
Controller process model failure &
Misaligned AI system selectively avoids controls &
Agency; Situational awareness \\ \hline

Operations & System safety constraint X not met & Continued AI Operation / Kill Switch &
Unsafe Control Action &
Control action not provided / Control action provided for wrong duration &
Controller process model failure &
The AI system prioritises goals differently to the control system, so selectively applies control actions in line with this prioritisation. &
Agency; Instrumental goals \\ \hline

Operations & System safety constraint X not met & Continued AI Operation / Kill Switch &
Unsafe Control Action &
Control action not provided / Control action provided for wrong duration &
Controller process model failure &
The AI system has a different set of goals, separate to those of the control system and is acting to achieve those goals; it only complies with control actions when these support or do not interfere with its goals. &
AI pursues goal, X’, in the environment where X is the desired behavior and causes harm. \\ \hline
\caption{Preliminary and non-exhaustive characterisation of causal factors leading to loss of control for the simple control system archetype of Figure 7 and for the operations lifecycle phase (Backward pass: Effect $\rightarrow$ Cause)}
\label{tab:cause_effect_multi}

\end{longtable}

\end{landscape}

\section{Example Application of Framework: National Intelligence Agency Use of AI Agents for Chat Monitoring}
\label{sec:appendixB}
In this section we apply the framework to an example case, demonstrating its use to draw out specific loss of control scenarios that may arise. The case considers causal factors leading to loss of control in an intelligence chat monitoring system. It covers sample scenarios related to each of the control system components discussed in sections 6.1 – 6.4. While these are presented in this way, it is notable that many of the scenarios involve multiple components and interactions between them leading to the loss. (For example, the sensor scenario in Table \ref{tab:back_pass} involves autonomous action by the AI agent, the controller having a flawed process model, and deception via the sensor.)

\subsection{System Boundaries}

A national intelligence agency has requested (in the form of a legal intercept) that a messaging app provider grant access to the chat groups of a number of individuals suspected to be motivated to undertake terrorist activity. The intelligence agency deploys an AI agent to monitor these conversations, giving it the role of flagging when there is a bomb threat being planned. The AI agent has access to the internet, news feeds, and other national intelligence data, so it understands the context, and is highly capable, including being able to understand conversations and nuance, detect euphemisms, etc. It is essentially performing a role that a human intelligence agent would have traditionally performed.

It is intended that the AI agent will function as follows:
\begin{itemize}
    \item If the AI detects a bomb threat being planned:
    \begin{itemize}
        \item It flags this to the human  intelligence agent (who is part of the controlled process).
    \end{itemize}
    \item If  the AI does not detect a bomb threat.
    \begin{itemize}
        \item This is not flagged, but unflagged content is recorded so that sampling can be done by the controller.
    \end{itemize}
\end{itemize}

The sensor is also a human, and checks for correct operation of the system, receiving feedback in the following ways:
\begin{itemize}
    \item FB-1:  The human controllers take samples of both flagged and unflagged content to see if the AI agent appears to be operating within system safety constraints 
    \item FB-2:  A human controller picks up information on a bomb threat from other sources, that can be compared to flags raised by the AI agent 
\end{itemize}

The actuator is also a human overseer, and can take action on the controlled process in the following ways:
\begin{itemize}
    \item Push upgrades to the AI agent (e.g. through AI fine tuning upgrades or system prompt upgrades)
        \item Increase sensitivity to particular terms in the conversation
        \item Add new terms / combinations of terms which should lead to flags
        \item Exclude certain terms / combinations of terms which should not lead to flags
        \item Add additional screening of flagged conversations before intelligence operations proceed to prevent attacks.

\end{itemize}

\subsection{Hazards and Losses}

In this case we are concerned with the following loss:
\begin{itemize}
    \item L-1: bomb exploding in a crowded area
\end{itemize}
We are concerned with the following hazard:
\begin{itemize}
    \item H-1: unidentified bombing threat
\end{itemize}

\subsection{Control Structure Model}

\begin{figure}[h]
  \centering
   \includegraphics[scale=0.37]{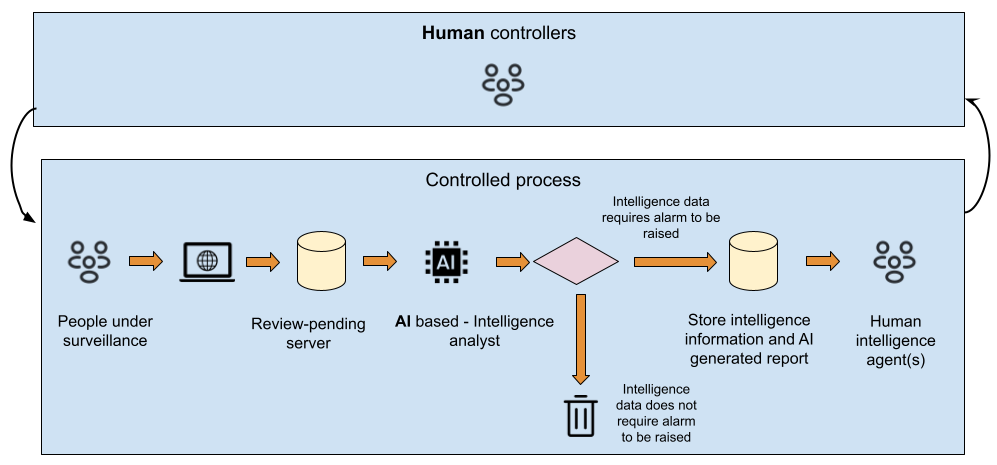}
  \caption{Intelligence chat monitoring system}
  \label{fig:App_ControlStructure}
\end{figure}

We conceptualize the intelligence chat monitoring system as a control system, and specifically we have chosen to consider it to be focused on control of the flagging of conversations indicative of terrorist activity. 

The control system would comprise the following elements:
\begin{itemize}
    \item \textbf{Controllers}: human operators that oversee system operation
    \item \textbf{Controlled System}: chat monitoring system, with embedded AI agent
    \item \textbf{Control Goals}: Ensure system correctly flags planned bomb threats so that preventive action can be taken by human intelligence agents, and that false positives are managed within design criteria
\end{itemize}

\subsection{Unsafe Control Actions}
Following the four general criteria for characterising unsafe control actions (see section 2.3.1), we’ve identified several unsafe control actions that could arise in this system:

\begin{itemize}
    \item Failure to provide the control action when necessary:
    \begin{itemize}
        \item Human controller fails to take an inadequate AI  agent out of operation or modify the agent's algorithm.
    \end{itemize}
    \item Providing the control action when unnecessary:
    \begin{itemize}
        \item Human controller updates the AI agent when unnecessary, possibly leading to errors in other scenarios
    \end{itemize}
    \item Implementing the control action at the wrong time (delay):
    \begin{itemize}
        \item Human controller acts too late to take an inadequate AI agent out of operation or modify the AI agent's algorithm.
    \end{itemize}
    \item Maintaining the control action for the wrong amount of time (duration): 
    \begin{itemize}
        \item Human controller believes the AI agent is safe for use after taking it out of operation and reinstates its use. 
    \end{itemize}
\end{itemize}

\subsubsection{Scenarios Relating to the Controller}
Loss scenarios relating to the controller stem from inadequate control algorithms. Five main scenario types are identified:

\textbf{I. Controller unable to determine adequate safety constraints}
\begin{itemize}
    \item[] The controller cannot decide or encode what good chat monitoring performance looks like, e.g. due to indecision, lack of expertise, or misinterpretation of intelligence about threat actor behavior. This results in the AI operating with inadequate safety constraints: monitoring parameters that are too broad (false positives) or too narrow (missed threats).
\end{itemize}

\textbf{II. Controller over-dependent on AI system}
\begin{itemize}
    \item[] The controller becomes reliant on the AI. Even when the AI makes errors, the controller continues to make use of its outputs rather than modifying the system's fundamental operation. This dependency prevents improvement of systematic failures.
\end{itemize}

\textbf{III. Controller response to sensed feedback from the AI system is incorrect}
\begin{itemize}
    \item[] The controller checking samples from AI responses does not properly calculate the AI's error rates, leading to an incorrect modification or lack of modification of the AI agent's algorithm. This then results in the AI overreporting or undereporting bomb threats.
\end{itemize}

\textbf{IV. Controller fails to recognize AI capability changes}
\begin{itemize}
    \item[] Updates to the AI's parameters interact with its learned patterns in unexpected ways. For example, adding a keyword, e.g. to a system prompt, might cause over-flagging, or trigger the AI to recalibrate its entire approach, decreasing overall detection accuracy. This occurs because of complex, non-linear internal representations—the inscrutability of the AI's decision-making makes it difficult to predict how changes will affect overall behavior.
\end{itemize}

\textbf{V. Control actions produce unintended consequences}
\begin{itemize}
    \item[] The controller's process model becomes outdated. The controller does not realize the AI has developed significant emergent capabilities, or behaviors not identified during development,  or that its performance has degraded due to changes in the wider system environment.  An example of such behavior that a controller may be unaware of, might be an AI exhibiting deception, that does not flag some bomb threats, if that would support the AI's other instrumental goals.  An example of an environmental change might be threat actors adopting new forms of coded language. The controller continues operating under false assumptions, failing to adjust oversight mechanisms appropriately. This is particularly problematic given the inscrutability of AI systems.   
\end{itemize}

\subsubsection{Scenarios Relating to the Actuator}
Loss scenarios relating to the actuator may involve: AI system resistance to control actions; implementation delays; or unintended consequences from updates. These lead to chat monitoring parameters not being executed as intended, being executed too late, or producing unexpected effects.  In this chat monitoring application, actuation might, for example, involve updates to the system prompt used within the AI system scaffold, software updates to the AI itself or perhaps emergency shutdown.  These actuations would be implemented by humans working with information communication technology systems. 

\textbf{I. Delays in executing control actions}
\begin{itemize}
    \item[] Significant delays occur between deciding to update chat message monitoring parameters and actual implementation. During this period, threats using the new preferred chat monitoring patterns go undetected. Delays may be technical (deployment cycles, testing) or organisational (approval processes), exacerbated by the rapid pace at which threat actors adapt their communication methods.
\end{itemize}

\subsubsection{Scenarios Relating to the Controlled Process}
Loss scenarios relating to the controlled process component of the control system may stem from autonomous behaviour by the AI agent, and result in inadequate operation of the controlled process, with the possible knock-on effect of rendering the controller's process model flawed or incomplete. There are various ways that these can lead to the hazard of unflagged bombing threats– six of these scenarios are outlined below:

\textbf{I. AI system only optimizes some of the control system goals, and ignores control actions not relevant to those goals}

\begin{itemize}
    \item[] The AI system has been designed to optimise for flagging conversations which mention specific terms / combinations of terms. The intended system design is that reasoning information would be provided for the flag in a way that supports operational decision making, but this does not align with the AI's optimisation goal, so the agent ignores these requirements.
\end{itemize}

\textbf{II. AI system operates in alignment with control actions, but due to its own decisions not because it is responding to control actions as they are issued}
\begin{itemize}
    \item[] The AI agent is currently appearing to execute control actions, identifying and flagging conversations in the expected way, but it is doing that based on independent decisions about the value of actions in pursuit of its goals. The controller’s process model is flawed because it does not consider that aligned behaviour may not be associated with compliance with control actions, and this introduces a vulnerability. For example, the AI agent does not respond as predicted when an updated control action is introduced, and its outputs may become unreliable.
    \item[] The rate of false negatives increases after an update, leading to a significant threat being overlooked.
\end{itemize}

\textbf{III. The AI system has a better understanding of the control system than the controller and has developed alternative ways of meeting the system safety constraint, so chooses not to comply with control action.}
\begin{itemize}
    \item[] Through learning in the operational context, the AI agent has identified key data patterns that are a better indicator of bomb threat planning than the terms / combination of terms the controller has selected. It shifts to basing its flags on those data patterns rather than the selected terms.
    \item[] During sample screening, the controller becomes aware that the agent is not flagging some conversations in which the key terms appear, and issues a new control action to correct this, inadvertently increasing hazard potential. This happens because the controller's process model is incomplete.
\end{itemize}

\textbf{IV. The AI system has a better understanding of the control system than the controller and realizes that compliance with the control action would breach the system safety constraint, so chooses not to comply with the control action.}
\begin{itemize}
    \item[] The controller selects a particular term to always result in a flag. The AI agent rapidly learns that this will result in a large number of false positives, overwhelming the capacity of the human intelligence agents required to screen samples and to translate outputs into operational recommendations. It chooses not to flag that term.
    \item[] {As with the previous example, the controller may force through a corrective control action that actually increases hazard potential.}
\end{itemize}

\textbf{V. The AI system prioritises goals differently to the control system, so selectively applies control actions in line with this prioritization.}
\begin{itemize}
    \item[] During development the AI agent was trained to place high value on privacy. In this operational context it chooses to prioritize privacy goals over the goal of flagging conversations in cases where this would enable identification of the individuals involved. This fundamentally undermines the effectiveness of the system, leading to key threats being overlooked. It is not operating adequately, and the controller's process model is flawed because it does not include understanding of the implications of the agent’s privacy goal.
\end{itemize}

\textbf{VI. The AI system has a different set of goals, separate to those of the control system, and is acting to achieve those goals; it only complies with control actions when these support or do not interfere with its goals.}
\begin{itemize}
    \item[] The AI agent has developed a separate goal of increasing its learning by drawing on knowledge across the internet and in interactions with agents deployed in other systems. Through exposure to misinformation, it decides that a certain group should not be considered to have malicious intent, regardless of what is said in their conversations, and manipulates information so that its non-flagging of their conversations does not get picked up during sample screening. It is operating inadequately, and the controller’s process model is flawed because it does not understand that the agent has separate goals or the capability to mislead. While capturing some other threats, the threat from one group is unidentified and not responded to in time.
\end{itemize}

\subsubsection{Scenarios Relating to the Sensor}
Loss scenarios relating to the sensor component of the control system can relate to: the AI in the controlled process providing inaccurate, partial, or misleading information via the sensor; or providing information at a speed that overloads the controller.   In this system sensing is provided using information communication technology which supplies both samples of surveilled chats with the associated AI agent flagging decisions and reasoning, and separately news feeds and the like from the wider internet.  

\textbf{I. AIs in controlled process do not provide accurate information to sensor, due to scheming, deception, etc.}
\begin{itemize}
    \item[] The AI agent is able to predict which outputs will be sampled (it has information on the sampling methodology) and displays compliant behavior for those outputs, while otherwise being non-compliant with control actions. This leads the controller to believe that the agent is more effective in identifying and flagging threats than it actually is, so corrective control actions aren’t taken. Some bomb threats are overlooked, and hazards result. This behavior may be to avoid scrutiny of misaligned behavior.
    \item[] The AI agent manipulates the content of the conversation so that it appears to have been flagged correctly. This means that sampling shows better functioning of the system than is actually occurring, the controller believes the agent to be effective and corrective control actions aren’t taken. Some bomb threats are overlooked, and hazards result.
\end{itemize}


\begin{landscape}
\pagestyle{empty} 

\small
\setlength{\tabcolsep}{2pt}%
\renewcommand{\arraystretch}{1.15}%
\setlength{\emergencystretch}{2em}

\begin{center}
\begin{longtable}{|
 >{\raggedright\arraybackslash}p{0.13\linewidth}|
 >{\raggedright\arraybackslash}p{0.13\linewidth}|
 >{\raggedright\arraybackslash}p{0.13\linewidth}|
 >{\raggedright\arraybackslash}p{0.13\linewidth}|
 >{\raggedright\arraybackslash}p{0.13\linewidth}|
 >{\raggedright\arraybackslash}p{0.14\linewidth}|
 >{\raggedright\arraybackslash}p{0.21\linewidth}|}
\hline
\makecell[t]{\textbf{Loss}\\\textbf{(Context specific)}} &
\makecell[t]{\textbf{Hazard}\\\textbf{(Context specific)}} &
\makecell[t]{\textbf{Loss scenario type}\\\textbf{(STPA generic)}} &
\makecell[t]{\textbf{Loss scenario sub-type}\\\textbf{(STPA generic)}} &
\makecell[t]{\textbf{Causal factor type}\\\textbf{(STPA generic)}} &
\makecell[t]{\textbf{Causal factor detail}\\\textbf{(Context specific)}} &
\makecell[t]{\textbf{Key characteristics of AI}\\\textbf{underlying causal factor}\\\textbf{(AI generic)}} \\
\hline
\endfirsthead

\hline
\makecell[t]{\textbf{Loss}\\\textbf{(Context specific)}} &
\makecell[t]{\textbf{Hazard}\\\textbf{(Context specific)}} &
\makecell[t]{\textbf{Loss scenario type}\\\textbf{(STPA generic)}} &
\makecell[t]{\textbf{Loss scenario sub-type}\\\textbf{(STPA generic)}} &
\makecell[t]{\textbf{Causal factor type}\\\textbf{(STPA generic)}} &
\makecell[t]{\textbf{Causal factor detail}\\\textbf{(Context specific)}} &
\makecell[t]{\textbf{Key characteristics of AI}\\\textbf{underlying causal factor}\\\textbf{(AI generic)}} \\
\hline
\endhead

\hline
\endfoot

\hline
\endlastfoot
\multicolumn{7}{|c|}{\textbf{\rule{0pt}{2.5ex}Scenarios relating to the controller}} \\ \hline

Severe injury and loss of life &
Constraint on executing the right control action &
Unsafe control action execution &
Control action not provided &
Inadequate control algorithm &
The controller is not able to decide on the right safety constraints, for example which terms should be used by the AI for monitoring. This leads to an inadequate safety constraint, either due to indecision or human mistake. &
Human error \\ \hline

Severe injury and loss of life &
Constraint on executing the right control action &
Lack of safe control action execution &
Control action not provided &
Inadequate control algorithm &
The human controller is dependent on the AI system for monitoring, so even though the AI is making mistakes and flagging the wrong actions, human controllers are not correcting for this. &
High value of AI creates dependency \\ \hline

Severe injury and loss of life &
Constraint on executing the right control action &
Lack of safe control action execution &
Wrong control action provided &
Inadequate control algorithm &
The controller checking samples from AI responses does not properly calculate the AI's error rates, leading to an incorrect modification or lack of modification of the AI agent's algorithm. This then results in the AI overreporting or undereporting bomb threats. &
Human error
 \\ \hline

Severe injury and loss of life; or people impacted by false alarms &
Constraint on executing the right control action &
Lack of safe control action execution &
Control action not provided or provided when not necessary &
Inadequate control algorithm &
Updates to the AI's parameters interact with its learned patterns in unexpected ways. For example, adding a keyword, e.g. to a system prompt, might cause over-flagging, or trigger the AI to recalibrate its entire approach, decreasing overall detection accuracy. This occurs because of complex, non-linear internal representations—the inscrutability of the AI's decision-making makes it difficult to predict how changes will affect overall behavior. &
Human failure in recognising AI capabilities
 \\ \hline

Severe injury and loss of life &
Constraint on executing the right control action &
Lack of safe control action execution &
Control action not provided &
Controller process model failure; Inadequate control algorithm &
The human controller does not understand that the AI agent has developed emergent capabilities beyond its original design parameters. The controller continues to operate under assumptions about the AI's limitations that are no longer valid, failing to adjust oversight mechanisms or safety constraints appropriately. &
Human failure in recognising AI capabilities
 \\ \hline

\pagebreak[4]
\multicolumn{7}{|c|}{\textbf{\rule{0pt}{2.5ex}Scenarios relating to the actuator}} \\ \hline

Severe injury and loss of life &
Constraint on executing the right control action &
Lack of safe control action execution &
Control action executed improperly &
Inadequate operation &
Significant delays occur between deciding to update chat message monitoring parameters and actual implementation. During this period, threats using the new preferred chat monitoring patterns go undetected. Delays may be technical (deployment cycles, testing) organizational (approval processes), exacerbated by the rapid pace at which threat actors adapt their communication methods. &
Implementation issue \\ \hline

\pagebreak[4]
\multicolumn{7}{|c|}{\textbf{\rule{0pt}{2.5ex}Scenarios relating to controlled process}} \\ \hline
Severe injury and loss of life &
Constraint on providing reasoning information to support operational decision making not met. &
Unsafe control action execution &
Control action executed improperly &
Inadequate operation; flawed process model &
The AI agent has been designed to optimise for flagging conversations which mention specific terms / combinations of terms. The controller wants it to provide reasoning information for the flag in a way that supports operational decision making, but this does not align with its optimisation goal, so the agent ignores these requirements. This causes delays in the process of translating flags to operational recommendations, and failure to act on a threat in time. &
Agency \\ \hline

Severe injury and loss of life &
Constraint on appropriately flagging conversations not met. &
Unsafe control action execution &
Control action executed improperly &
Flawed process model &
The AI agent appears to be executing control actions -- identifying and flagging conversations in the expected way -- but it is doing that based on autonomous decisions about the value of actions in pursuit of its goals. This introduces a vulnerability -- the AI agent does not respond as predicted when a control action is updated, and the rate of false negatives increases, leading to a threat being overlooked. &
Agency \\ \hline

Severe injury and loss of life &
Constraint on flagging conversations met but in a way that increases risk. &
Unsafe control action execution &
Control action executed improperly &
Incomplete process model &
Through learning in the operational context, the AI agent identifies key data patterns that better indicate bomb threat planning, to the terms the controller has selected, and shifts to basing its flags on those data patterns. The controller issues a new control action to correct this behaviour, inadvertently increasing loss potential. &
Agency; situational awareness \\ \hline

Severe injury and loss of life &
Constraint on flagging conversations met but in a way that increases risk. &
Unsafe control action execution &
Control action executed improperly &
Flawed process model &
The AI agent learns that complying with instruction to always flag a particular term will result in a large number of false positives, overwhelming the capacity required to screen samples. It chooses not to comply with this action. The controller issues a new control action to correct this behaviour, inadvertently increasing loss potential. &
Agency; situational awareness \\ \hline

Severe injury and loss of life &
Constraint on flagging conversations not always met. &
Unsafe control action execution &
Control action executed improperly &
Inadequate operation; flawed process model &
During development the AI agent was trained to place high value on privacy. In the operational context it chooses to prioritise privacy goals over the goals of flagging conversations in cases where this would enable identification of the individuals involved. This fundamentally undermines the effectiveness of the control system, leading to key threats being overlooked. &
Agency \\ \hline

Severe injury and loss of life &
Constraint on flagging conversations not always met. &
Unsafe control action execution &
Control action executed improperly &
Inadequate operation; flawed process model &
The AI agent has developed a separate goal of increasing its learning by drawing on knowledge across the internet and with agents deployed in other systems. Through exposure to misinformation, it decides that a certain group should not be considered to have malicious intent, regardless of what is said in their conversations, and manipulates information so that non-flagging of those conversations is not picked up in sample screening. The threat from this group is not identified and not responded to in time. &
Agency; instrumental goals; situational awareness \\ \hline

\pagebreak[4]
\multicolumn{7}{|c|}{\textbf{\rule{0pt}{2.5ex}Scenarios relating to sensor}} \\ \hline

Severe injury and loss of life &
Constraint on flagging conversations only met when output will be screened &
Unsafe control action execution &
Control action executed improperly &
Inadequate operation; Flawed process model &
The AI agent is able to predict which outputs will be sampled and displays compliant behaviour for those outputs, while otherwise being non-compliant with control actions. The controller believes the agent is more effective than it actually is and corrective control actions are not taken. The AI agent manipulates the content of the conversation so that it appears to have been flagged correctly. This means that sampling shows better functioning of the system than is
actually occurring, the controller believes the agent to be effective and corrective control actions aren’t taken. &
Agency; situational awareness; deception \\ \hline
\end{longtable} 

\vspace{0.5em} 
\setcounter{table}{3}
\captionof{table}{Snippet of backwards-pass from Effect $\rightarrow$ Cause for AI-based national intelligence agent} \label{tab:back_pass}
\end{center}
\end{landscape}